\documentclass[aps,prd,preprint,superscriptaddress,amssymb,amsmath,nofootinbib]{revtex4}

\usepackage{color}
\usepackage{epsfig}
\usepackage{graphicx}
\usepackage{float}
\usepackage{epstopdf}
\usepackage{hyperref} 
\usepackage[usenames,dvipsnames]{xcolor}
\usepackage{multirow}
\usepackage[normalem]{ulem}
\usepackage{slashed}
\usepackage{feynmf}

\begin{document}

\preprint{}

\title{Effective Field Theory of Chirally-Enhanced Muon Mass and Dipole Operators}

\author{Radovan Dermisek}
\email[]{dermisek@indiana.edu}
\affiliation{Department of Physics, Indiana University, Bloomington, IN, 47405, USA}

\author{Keith Hermanek}
\email[]{khermane@iu.edu}
\affiliation{Department of Physics, Indiana University, Bloomington, IN, 47405, USA}

\author{Navin McGinnis}
\email[]{nmcginnis@triumf.ca}
\affiliation{TRIUMF, 4004 Westbrook Mall, Vancouver, BC, Canada V6T 2A3}

\author{Sangsik Yoon}
\email[]{yoon12@iu.edu}
\affiliation{Department of Physics, Indiana University, Bloomington, IN, 47405, USA}

\date{May 1, 2023}

\begin{abstract} 
We study corrections to observables related to the muon in the context of models of new physics which generate mass-enhanced corrections to the muon dipole moments. Working in the Standard Model effective theory, we demonstrate a correlation between the decay of the Higgs boson to muons, and the magnetic and electric dipole moments of the muon generated by the dominant matching corrections. This defines a novel way to classify predictions for a wide variety of models of new physics based on the pattern of deviations of these three observables. In particular, when applied to specific models we find that this correlation has a potential to rule out whole models or set upper bounds on the scale of new physics motivated by the muon anomalous magnetic moment.
\end{abstract}


\maketitle

\section{Introduction}
The Standard Model (SM) has been established as the consistent description of particle physics at the electroweak (EW) scale. In particular, the Higgs mechanism provides the necessary ingredient to understand the origin of masses of all known fundamental particles. The precision of this picture is currently being tested in large part at the Large Hadron Collider (LHC) and will guide the ongoing efforts to understand in greater detail the role of the Higgs boson in nature and its possible connection to physics beyond the SM.

Although several couplings of the SM Higgs are well understood from an experimental standpoint~\cite{ATLAS:2022vkf,CMS:2022dwd}, the current precision of its couplings to the second generation fermions remains fairly lacking. In particular the decay of the Higgs boson to muon pairs may deviate more than twice from what is expected in the SM~\cite{ATLAS:2020fzp}. On the other hand, the muon anomalous magnetic moment has raised an intriguing puzzle, where its recently measured value was found to be more than four standard deviations away from the SM prediction~\cite{Muong-2:2021ojo,Aoyama:2020ynm}. While efforts to improve the theoretical prediction are still underway~\cite{Borsanyi:2020mff,CMD-3:2023alj}, the latter observation has garnered much attention in the literature as a possible hint for new particles in nature. The possibilities range from new gauge forces, fermionic matter, or extensions of the Higgs sector, see Refs.~\cite{Czarnecki:2001pv,Freitas:2014pua,Lindner:2016bgg} and references thererin.

Among the proposed explanations for the muon anomalous magnetic moment puzzle, those which lead to a \textit{chiral enhancement} in the muon dipole operator are associated with scenarios with the largest mass scales of new physics~\cite{Kannike:2011ng,Dermisek:2013gta,Capdevilla:2020qel,Capdevilla:2021rwo,Crivellin:2021rbq,Stockinger:2022ata,Crivellin:2022wzw,Dermisek:2022aec}. In these models, the typical contribution to the muon magnetic moment scales as $\Delta a_{\mu}\simeq m_{\mu}\lambda^{3}v/16\pi^{2}M^{2}$ where $\lambda^{3}$ and $M^{2}$ are representative of products of individual, fundamental couplings and masses of new particles, whereas the naive scaling due to contributions from new particles would be $\Delta a_{\mu}\simeq \lambda^{2}m_{\mu}^{2}/16\pi^{2}M^{2}$. Thus, we see that the enhancement appears as a factor of $\sim \lambda v/m_{\mu}$ compared to the naive scaling. Hints of such models may also exist in $B$ anomalies~\cite{Raby:2017igl,Crivellin:2018qmi,Barman:2018jhz,Arnan:2019uhr,1906.11297}, models of radiative lepton masses~\cite{Yin:2021yqy,Thalapillil:2014kya}, the recently studied Cabibbo anomaly~\cite{Endo:2020tkb,Crivellin:2020ebi,Crivellin:2022ctt}, Higgs decays~\cite{Crivellin:2020tsz,Crivellin:2021rbq}, and physics of the dark sector~\cite{Kowalska:2017iqv,Calibbi:2018rzv,Kowalska:2020zve,Jana:2020joi,Athron:2021iuf,Arcadi:2021cwg,Cai:2021nmk}. However, a drawback to these scenarios is that increasingly large scales will be difficult or even impossible to test at the LHC and future colliders. We demonstrate that models which generate a chirally-enhanced contribution to the muon dipole moment through the coupling of the Higgs boson to new particles simultaneously generate a modification of the Higgs coupling to the muon which could be observed through the corresponding modification of the $h\to\mu^{+}\mu^{-}$ branching ratio. Further, we elaborate on our recent proposal to correlate these observations with future measurements of the electric dipole moment of the muon~\cite{Dermisek:2022aec}. In this respect, we argue that while direct evidence for new physics may be out of reach for colliders in these scenarios, the pattern of deviations of SM couplings in the low-energy effective field theory is distinct and therefore sharp conclusions can still be made.

We discuss models where new particles are allowed to couple to the muon through the SM Higgs boson at tree- or one-loop level. Simplified models with loop-level mixing have been extensively studied previously~\cite{Capdevilla:2020qel,Capdevilla:2021rwo,Crivellin:2021rbq}. To generalize previous observations and encompass all relevant models in a unified way, we outline our arguments using the SM effective field theory (SMEFT) and provide model independent relations. Using this machinery we show how, in specific models, the correlation of the Higgs decay to muon dipole moments can be used to set an upper bound on the scale of new physics which can be stronger than that from more general constraints from perturbative unitarity or other unphysical regions of parameter space.

This paper is organized as follows. In Section~\ref{sec:smeft_ops}, we discuss the SMEFT operators relevant for our main results and outline our notation. In Section~\ref{sec:models}, we outline the matching of individual models in SMEFT showing, in particular, the predicted correlations of operators connecting Higgs decays to the muon dipole moments. In Section~\ref{sec:ellipse}, we present our main results and discuss implications for upcoming precision measurements related to the muon followed by a discussion of non-minimal models in Section~\ref{sec:discussion}. We conclude in Section~\ref{sec:conclusions}. We also provide detailed appendices useful for approximate formulas appearing in the text.

%
%
\section{Effective interactions}
\label{sec:smeft_ops}
We focus on the dimension-six operators in SMEFT which generate modifications of the muon coupling to the Higgs as well as muon dipole moments. In a given basis, these effects are captured by operators coupling left- and right-handed muon fields, $\bar{l}_{L}\mathcal{O}e_{R}$. In the Warsaw basis~\cite{Grzadkowski:2010es}, and including the tree-level muon Yukawa coupling, the effective lagrangian relevant to this discussion comprises of four operators
\begin{flalign}
-\mathcal{L}\supset y_{\mu}&\bar{l}_{L}\mu_{R}H + C_{\mu H}\bar{l}_{L}\mu_{R}H\left(H^{\dagger}H\right)+C_{\mu W}\bar{l}_{L}\sigma^{\mu\nu}\mu_{R}\tau^{I}HW^{I}_{\mu\nu}+C_{\mu B}\bar{l}_{L}\sigma^{\mu\nu}\mu_{R}HB_{\mu\nu} + h.c.,
\label{eq:eff_lagrangian}
\end{flalign}
where the doublet components of the muon field are labeled as $l_{L}=(\nu_{\mu}, \mu_{L})^{T}$, $\tau^{I}$ are the Pauli matrices, and the gauge field strength tensors are given by
\begin{flalign}
W^{I}_{\mu\nu}=&\partial_{\mu}W_{\nu} - \partial_{\nu}W_{\mu}-g\varepsilon^{IJK}W^{J}_{\mu}W^{K}_{\nu}\\
B_{\mu\nu}=&\partial_{\mu}B_{\nu} - \partial_{\nu}B_{\mu}.
\end{flalign}
We assume that all parameters in Eq.~\ref{eq:eff_lagrangian} can be complex. Other operators sharing the chiral structure $\bar{l}_{L}\mathcal{O}e_{R}$ include only a few cases of four-fermion operators. These operators are relevant to the one-loop renormalization of Eq.~\ref{eq:eff_lagrangian}. For the moment we will restrict our discussion to the tree-level predictions of Eq.~\ref{eq:eff_lagrangian}, and defer a discussion of renormalization group (RG) effects to later sections.

When the Higgs develops a vev
\begin{equation}
H=\begin{pmatrix}0\\
		v+\frac{1}{\sqrt{2}}h
	\end{pmatrix},\;\;\;\; v=174\text{ GeV},
\end{equation}
triggering EWSB, the first two terms in Eq.~\ref{eq:eff_lagrangian} generate the mass and Higgs coupling to the muon. Written in terms of Dirac spinors, we have
\begin{equation}
-\mathcal{L}\supset m_{\mu}\bar{\mu}\mu + \frac{1}{\sqrt{2}}\left(\lambda^{h}_{\mu\mu}\bar{\mu}P_{R}\mu h + h.c.\right),
\label{eq:mass_lag}
\end{equation}
where $m_{\mu}$ is the physical muon mass and
\begin{flalign}
m_{\mu}=&\left(y_{\mu}v + C_{\mu H}v^{3}\right)e^{-i\phi_{m_{\mu}}},\label{eq:mmu}\\
\lambda_{\mu\mu}^{h}=&\left(y_{\mu} + 3C_{\mu H}v^{2}\right)e^{-i\phi_{m_{\mu}}}.
\label{eq:mmu_lamhmu}
\end{flalign}
The overall phase, $\phi_{m_{\mu}}$, appears through a redefinition of the muon fields, $\bar{\mu}_{L}\mu_{R}\to e^{-i\phi_\mu}\bar{\mu}_{L}\mu_{R}$, to make the mass term real and positive. 

Additional corrections to the Higgs coupling to the muon arise from the dimension-six operators $C_{H\Box}(H^{\dagger}H)\Box(H^{\dagger}H)$ or $C_{HD}(H^{\dagger}D^{\mu}H)^{*}(H^{\dagger}D_{\mu}H)$. After EWSB, both operators result in non-canonical corrections to the physical Higgs kinetic term. This leads to a universal shift of all Higgs couplings to SM fermions, and in particular $y_{\mu}\rightarrow y_{\mu}(1+C_{HD}v^{2}-C_{H\Box}v^{2}/4)$ in Eq~\ref{eq:mmu_lamhmu}, see~\cite{Alonso:2013hga}. Thus, the contribution of these operators to the Higgs coupling to the muon propagates as corrections suppressed by $\sim m_{\mu}/v$ compared to that of $C_{\mu H}$. This suppression is in principle compensated in a given model if it happens that $C_{HD}$ and $C_{H\Box}$ are generated at tree-level at the matching scale while $C_{\mu H}$ is generated at one loop, e.g. as in the SM extended with a singlet scalar~\cite{Jiang:2018pbd}. For the models we focus on in the following sections $C_{HD}$ and $C_{H\Box}$ are always generated at one loop. Thus, we will ignore these operators in our main discussion as they will be suppressed by a loop factor in addition to power counting in $m_{\mu}/v$.

Due to the different combinatorial factor accompanying the corrections proportional to $C_{\mu H}$ in Eqs.~\ref{eq:mmu} and~\ref{eq:mmu_lamhmu}, the Higgs coupling to the muon is necessarily modified compared to that in the SM, $(\lambda^{h}_{\mu\mu})_{SM}=m_{\mu}/v$. In the basis where the muon mass is real and positive, we define
\begin{equation}
R_{h\to\mu^{+}\mu^{-}}  \equiv \frac{BR(h\to\mu^{+}\mu^{-})}{BR(h\to\mu^{+}\mu^{-})_{SM}}=\left(\frac{v}{m_{\mu}}\right)^{2}\big|\lambda_{\mu\mu}^{h}e^{-i\phi_{m_{\mu}}}\big|^{2},
\end{equation}
for which the most up-to-date measurements of $h\to\mu\mu$ set an upper limit of $R_{h\to\mu^{+}\mu^{-}} \leq2.2$~\cite{ATLAS:2020fzp,CMS:2020xwi}. Unless the modification to $(\lambda^{h}_{\mu\mu})_{SM}$ is only a pure phase, we see that the decay rate is necessarily modified.

Using Eqs~\ref{eq:mmu} and \ref{eq:mmu_lamhmu} gives
\begin{flalign}
R_{h\to\mu^{+}\mu^{-}} = 1 + 4\textrm{Re}\left(\frac{C_{\mu H}v^{3}}{m_{\mu}}e^{-i\phi_{m_{\mu}}}\right) + 4\left[\textrm{Re}\left(\frac{C_{\mu H}v^{3}}{m_{\mu}}e^{-i\phi_{m_{\mu}}}\right)\right]^{2}+ 4\left[\textrm{Im}\left(\frac{C_{\mu H}v^{3}}{m_{\mu}}e^{-i\phi_{m_{\mu}}}\right)\right]^{2}.
\label{eq:R_mu_def}
\end{flalign}
We note that corrections to the muon mass proportional to $\textrm{Re}(C_{\mu H})$ can deviate $R_{h\to\mu^{+}\mu^{-}}$ away from 1 in either direction depending on the sign of $\textrm{Re}(C_{\mu H})$, whereas corrections proportional to $\textrm{Im}(C_{\mu H})$ can only enhance $R_{h\to\mu^{+}\mu^{-}}$.

After EWSB the second two terms in Eq.~\ref{eq:eff_lagrangian} combine to generate the electric and magnetic dipole moment of the muon
\begin{equation}
\mathcal{L}\supset -C_{\mu \gamma}ve^{-i\phi_{m_{\mu}}}\bar{\mu}_{L}\sigma^{\mu\nu}\mu_{R}F_{\mu\nu} + h.c.
\end{equation}
with $C_{\mu\gamma}=(C_{\mu B}c_{W}-C_{\mu W}s_{W})$, where $s_{W}\equiv \sin\theta_{W}$ is the Weinberg angle. Defining the electric and magnetic dipole moments in terms of Dirac spinors, in the basis where the muon mass is real and positive,
\begin{equation}
\mathcal{L}\supset \frac{\Delta a_{\mu}e}{4m_{\mu}}\bar{\mu}\sigma^{\mu\nu}\mu F_{\mu\nu} - \frac{i}{2}d_{\mu}\bar{\mu}\sigma^{\mu\nu}\gamma^{5}\mu F_{\mu\nu},
\label{eq:ldipole}
\end{equation}
we have that 
\begin{flalign}
\Delta a_{\mu} &=-\frac{4m_{\mu}v}{e}\textrm{Re}[C_{\mu \gamma}e^{-i\phi_{m_{\mu}}}]\label{eq:mdipole}\\
d_{\mu} &= 2v\textrm{Im}[C_{\mu \gamma}e^{-i\phi_{m_{\mu}}}].
\label{eq:edipole}
\end{flalign}
Our convention in defining the dipole moments is such that the electromagnetic (EM) charge unit $e>0$.\;\footnote{It is worth noting the sign conventions appearing in Eqs.~\ref{eq:ldipole},~\ref{eq:mdipole}, and \ref{eq:edipole}. The signs are chosen so that the definition of dipole moments from the Lagrangian, Eq.~\ref{eq:ldipole}, matches that typically used in the literature, accounting for both our convention for the sign of $e$ and using the mostly-minus metric $\eta=\text{diag}(+1,-1,-1,-1)$.} The recent measurement of the muon anomalous magnetic moment provides~\cite{Muong-2:2021ojo,Aoyama:2020ynm}
\begin{equation}
\Delta a_{\mu}=(2.51\pm 0.59)\times 10^{-9}.
\end{equation}
For the electric dipole moment, there is currently both a direct limit from the Brookhaven Muon $g-2$ results~\cite{Muong-2:2008ebm}:
\begin{equation}
|d_{\mu}|<1.9\times 10^{-19}\; [e\cdot cm],
\end{equation}
as well as an indirect limits based on the Schiff moments of heavy molecules~\cite{Ema:2021jds}:
\begin{equation}
|d_{\mu}| < 1.9\times 10^{-20}\; [e\cdot cm ],
\end{equation}
where we have quoted the more stringent bound based on ThO. Currently there are two new proposals to improve this limit. The Fermilab (FNAL) Muon $g-2$ collaboration projects~\cite{Lukicov:2019ibv} an improvement down to
\begin{equation}
|d_{\mu}| < 10^{-21}\; [e\cdot cm],
\end{equation}
while a new experiment based on frozen-spin technique, proposed to be hosted at the Paul Scherrer Institute (PSI) projects \cite{Adelmann:2021udj}
\begin{equation}
|d_{\mu}| < 6\times 10^{-23}\; [e\cdot cm].
\end{equation}
Other operators involving a single leptonic current of the muon fields, such as $C^{(1)}_{Hl}H^{\dagger}i\overset{\leftrightarrow}{D_{\mu}}H\left(\bar{l}_{L}\gamma^{\mu}l_{L}\right)$, $C^{(3)}_{Hl}H^{\dagger}i\overset{\leftrightarrow}{D_{\mu}^{I}}H\left(\bar{l}_{L}\tau^{I}\gamma^{\mu}l_{L}\right)$, and $C_{He}H^{\dagger}i\overset{\leftrightarrow}{D_{\mu}}H\left(\bar{\mu}_{R}\gamma^{\mu}\mu_{R}\right)$ lead to corrections of the muon couplings to gauge bosons. While the corresponding Wilson coefficients are not central to the main observables that we discuss, they are subject to constraints for a given matching scale, $\Lambda$, as they lead to corrections to electroweak precision observables (EWPO) such as the partial width of the $Z$-boson and the muon lifetime~\cite{Kannike:2011ng,Dermisek:2021ajd}. See also the appendix of Ref.~\cite{Crivellin:2021rbq} for general expressions of couplings for the $Z$- and $W$-bosons to the muon in terms of $C^{(1)}_{Hl}$, $C^{(3)}_{Hl}$, and $C_{He}$.

In the following sections, we will argue that while the observables $\Delta a_{\mu}$, $d_{\mu}$, and $R_{h\to\mu^{+}\mu^{-}}$ need not be related in general, they are tightly correlated in models of new physics which aim to explain $\Delta a_{\mu}$ via a mass-enhanced correction.

\section{Mass enhanced corrections}
\label{sec:models}
Despite the proliferation of effective operators it is often the case that the effective field theory simplifies in the context of concrete models, where subsets of operators may be dictated by the same couplings if they are even generated at all. In this section, we outline two classes of models which are known to generate mass enhanced corrections to $\Delta a_{\mu}$ and discuss their matching onto the effective theory. The two classes of models are distinguished depending on whether the dominant contribution to $C_{\mu H }$ is generated at tree- or one-loop level at the matching scale and refer to these cases hereafter as \textit{tree models} and \textit{loop models}, respectively. The tree models have been enumerated and studied in connection with $\Delta a_{\mu}$ in~\cite{Kannike:2011ng,Dermisek:2013gta}, whereas similar studies for loop models can be found in~\cite{Calibbi:2018rzv,Crivellin:2018qmi,Crivellin:2021rbq}. Further, a new class of models which generate chirally-enhanced corrections to $\Delta a_{\mu}$ has recently been identified~\cite{Guedes:2022cfy}, which we will refer to as \textit{bridge models}. In~\cite{Dermisek:2022aec}, we argued that $C_{\mu H }$  and $C_{\mu\gamma}$ are sharply correlated at the matching scale in the tree and loop models. Here, we elaborate on this relationship and expand the scope of our arguments to a subset of the bridge models.
%
%
\subsection{Tree models}
\begin{figure}[t]
\includegraphics[width=1\linewidth]{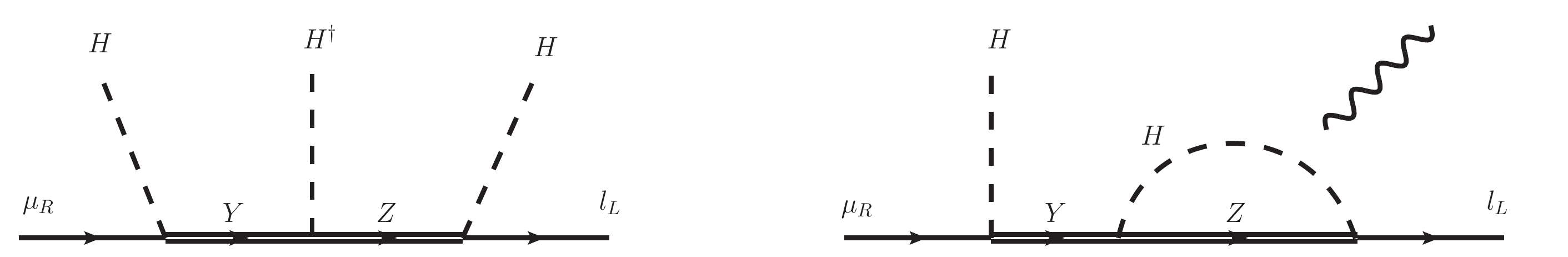}
\caption{A representative tree-level diagram generating $C_{\mu H}$ in models with heavy leptons which couple to the muon through the SM Higgs (left), and a corresponding diagram leading to the mass-enhanced correction to $\Delta a_{\mu}$.}
\label{fig:tree_diags}
\end{figure}

Models where $C_{\mu H}$ is generated at tree-level at the matching scale consists of UV completions with two new fermion fields which have a mutual coupling to the SM Higgs boson and tree level mixing with the left- and right-handed muon fields. In the left panel of Fig.~\ref{fig:tree_diags}, we show a representative Feynman diagram which generates $C_{\mu H}$ when the heavy fields, $Y$ and $Z$, are integrated out. The Higgs legs on the diagram are generically labeled as $H$. However, gauge invariance will force one leg to be $H^\dagger$ once quantum numbers of the new leptons are chosen. The direction of charge arrows on the internal fermion lines will similarly be enforced in a given model. We consider models where $Y$ and $Z$ are vectorlike leptons and thus have contributions to their mass which do not originate from EWSB. 

Starting from the left diagram in Fig.~\ref{fig:tree_diags}, we see that by connecting $H^{\dagger}H$ pairs of the external Higgs legs and dressing the resulting diagram by all possible insertions of a photon leg constructs the contributions to $C_{\mu\gamma}$, as in the right panel of Fig.~\ref{fig:tree_diags}, which lead to mass-enhanced corrections to $\Delta a_{\mu}$.~\footnote{Note that by replacing $Y\to l_{L}$ or $Z\to \mu_{R}$ in Fig.~\ref{fig:tree_diags} (left) leads to a tree-level correction to $C_{\mu H}$ which is proportional to the muon Yukawa coupling. We neglect these corrections in the tree models as they are $m_{\mu}/v$ suppressed compared to the corrections we discuss.} Thus, in the tree models it is expected that the same couplings needed to generate $C_{\mu H}$ simultaneously give a contribution to $C_{\mu\gamma}$, and the two Wilson coefficients will be directly related without a free parameter. Indeed, as was studied in~\cite{Kannike:2011ng,Dermisek:2021ajd,Dermisek:2022aec}, the dominant contributions to $C_{\mu H}$ and $C_{\mu\gamma}$ are related and lead to the following correlation between Wilson coefficients at the matching scale
\begin{equation}
C_{\mu\gamma} \simeq \frac{\mathcal{Q}e}{64\pi^{2}}C_{\mu H},
\label{eq:tree_formula}
\end{equation}
where $\mathcal{Q}$ is an integer factor depending only on the quantum numbers of the new leptons. In Table~\ref{table:models}, we list the quantum numbers under $SU(2)_{L}\times U(1)_{Y}$ of the possible pairs of new leptons in tree models, and the corresponding $\mathcal{Q}$-factor relating the Wilson coefficients via Eq.~\ref{eq:tree_formula}. Since these models require couplings of heavy fermions to both the muon and SM Higgs fields to generate these corrections, the possible models are highly limited by the allowed quantum numbers.
\begin{table}[t]
\begin{center}
\begin{tabular}{ |c||c| } 
\hline
$SU(2)\times U(1)_{Y}$& $\mathcal{Q}$ \\
\hline
$\mathbf{2}_{-1/2}\oplus\mathbf{1}_{-1}$ & 1  \\
\hline
$\mathbf{2}_{-1/2}\oplus\mathbf{3}_{-1}$  & 9 \\
\hline
$\mathbf{2}_{-3/2}\oplus\mathbf{1}_{-1}$ & 5 \\
\hline
$\mathbf{2}_{-3/2}\oplus\mathbf{3}_{-1}$ & 5 \\
\hline
$\mathbf{2}_{-1/2}\oplus\;\mathbf{3}_{0}$ &1 \\
\hline
\end{tabular}
\caption{Quantum numbers of $Y\oplus Z$ fields under $SU(2)\times U(1)_{Y}$ and corresponding $\mathcal{Q}$-factor relating $C_{\mu\gamma}$ and $C_{\mu H}$ in the five tree models~\cite{Kannike:2011ng}. The hypercharge numbers are normalized so that $Q_{EM}=T^{3}+Y$.}
\label{table:models}
\end{center}
\end{table}

In~\cite{Kannike:2011ng,Dermisek:2021ajd,Dermisek:2022aec}, Eq.~\ref{eq:tree_formula} was presented after lengthy calculations in the mass eigenstate basis involving loops of EW gauge bosons in addition to Higgs mediated diagrams.  To demonstrate that this correlation appears in tree models at the matching scale via the diagramatic arguments we have just presented we consider as a specific case the SM extended with a vectorlike doublet and charged singlet leptons, $L_{L}$ and $E_{R}$, whose quantum numbers mirror those of the left- and right-handed muon fields (corresponding to the first row in Table~\ref{table:models}). The most general lagrangian of Yukawa and mass terms is then
\begin{flalign}\nonumber
-\mathcal{L}\supset \;&y_{\mu}\bar{l}_{L}\mu_{R}H + \lambda_{L}\bar{L}_{L}\mu_{R}H + \lambda_{E}\bar{l}_{L}E_{R}H + \lambda\bar{L}_{L}E_{R}H + \bar{\lambda}H^{\dagger}\bar{E}_{L}L_{R}\\
&+M_{L}\bar{L}_{L}L_{R}+M_{E}\bar{E}_{L}E_{R} + h.c.,
\end{flalign}
where the doublet components are labeled as $L_{L,R}=(L^{0}_{L,R},L^{-}_{L,R})^{T}$. We work in the unbroken phase of the SM where the full $SU(2)\times U(1)_{Y}$ is linearly realized. Integrating out heavy leptons at tree level gives
\begin{figure}[t]
\includegraphics[width=1\linewidth]{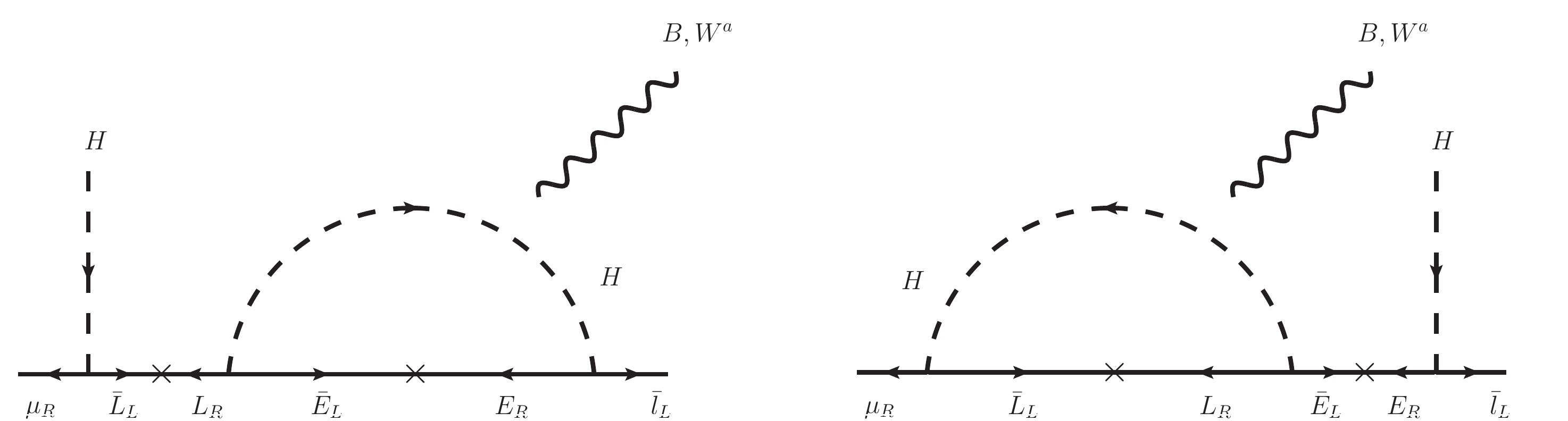}
\caption{Dominant contributions to $C_{\mu B,W}$ in the $\mathbf{2}_{-1/2}\oplus\mathbf{1}_{-1}$ tree model. The $B,W^{a}$ fields are attached to all particles in the loop. In these diagrams, arrows on fermion lines denote chiral flow.}
\label{fig:Cmugamma}
\end{figure}
\begin{equation}
C_{\mu H}=\frac{\lambda_{L}\lambda_{E}\bar{\lambda}}{M_{L}M_{E}}.
\end{equation}
To calculate the chirally-enhanced contributions to $\Delta a_{\mu}$ in this model we consider the diagrams constructed as we described by connecting $H^{\dagger}H$ pairs in all possible ways and dressing the diagram with all possible insertions of the $B_{\mu}$ and $W^{a}_{\mu}$ gauge fields, shown in Fig.~\ref{fig:Cmugamma}. We note that corrections obtained via an insertion of the $B_{\mu}$ and $W^{a}_{\mu}$ on the internal heavy fermion propagator which is not in the loop result in renormalization of the respective gauge charges. Thus, we need only compute diagrams constructed from $B_{\mu}$ and $W^{a}_{\mu}$ insertions for particles in the loop. Since we are interested in corrections when $M_{L,E}\gg v$, propagators of $L_{L,R}$ and $E_{L,R}$ are treated in the heavy mass limit. For the dipole operators, we find
\begin{flalign}
C_{\mu B}=&-\frac{g^{\prime}}{64\pi^{2}}\frac{\lambda_{L}\lambda_{E}\bar{\lambda}}{M_{L}M_{E}}\left[\sum_{j=L,R}\left(Y_{E_{j}}F(x_{E})+2Y_{L_{j}}F(x_{L})\right)+Y_{H}\left(2G(x_{L}) -G(x_{E})\right)\right]\\
C_{\mu W}=&\frac{g}{128\pi^{2}}\frac{\lambda_{L}\lambda_{E}\bar{\lambda}}{M_{L}M_{E}}G(x_{E}),
\end{flalign}
where $Y_{L_{L,R}}$, $Y_{E_{L,R}}$, and $Y_{H}$ are the $U(1)_{Y}$ charges of the heavy lepton doublets, singlets, and SM Higgs doublet, respectively,
\begin{flalign}
F(x)&=-\frac{x^{3}-4x^{2}+3x+2x\ln(x)}{2(1-x)^{3}},\label{eq:Ffunc}\\
G(x)&=\frac{x-x^{3}+2x^{2}\ln(x)}{(1-x)^{3}},\label{eq:Gfunc}
\end{flalign}
and $x_{L,E}=M^{2}_{L,E}/m_{h}^{2}$.
After EWSB, and identifying the factor corresponding to $C_{\mu H}$ we have that
\begin{equation}
C_{\mu\gamma}=-\frac{e}{64\pi^{2}}\left[\sum_{j=L,R}\sum_{i=L,E}Q_{i_{j}}F(x_{i}) + Q_{G^{+}}G(x_{L})\right]C_{\mu H},
\end{equation}
where $Q_{L_{L,R},E_{L,R}}=-1$ and $Q_{G^{+}}=+1$ are the EM charges of the charged components of all new fermion degrees of freedom and charged Goldstone in the SM Higgs doublet, respectively. Note that we have performed the calculation in the Feynman gauge where the Goldstones appear as massive particles and we have approximated the Goldstone loops assuming $m_{G^{0}}=m_{G^{\pm}}=m_{h}$.
Taking the limit $x_{i}\gg 1$, $F(x_{i})\to 1/2$, $G(x_{i})\to 1$, we finally obtain
\begin{equation}
C_{\mu \gamma}\simeq\frac{e}{64\pi^{2}}C_{\mu H},
\end{equation}
which reproduces the known result quoted in Table~\ref{table:models} which was originally calculated in the mass eigenstate basis.

The calculation for other models in Table~\ref{table:models} proceeds similarly, and thus for tree models the generic correlation between $C_{\mu H}$ and the dipole operator can be parameterized as
\begin{equation}
C_{\mu\gamma}\simeq-\frac{e}{64\pi^{2}}C_{\mu H}\sum_{j=L,R}\sum_{i=\text{charged fermions}}Q_{i_{j}}a_{i_{j}}(x_{L},x_{E}) + Q_{G^{+}}b(x_{L},x_{E})
\label{eq:tree_gen}
\end{equation}
where $a(x_{L},x_{E})$ are real numbers parameterizing loops with heavy fermions and the physical Higgs and $b(x_{L},x_{E})$ parameterizes the Goldstone-mediated contributions. In Table~\ref{table:models_loops}, we list all particles charged under $Q_{EM}$ in each of the tree models and their corresponding contribution to the sum in Eq.~\ref{eq:tree_gen}. The $F(x)$ and $G(x)$ loop functions are as defined above, whereas the remaining functions needed to complete the table are given by
\begin{flalign}
A(x,y)&=\frac{xy}{2}\left(\frac{-3y+x(1+x+y)}{(1-x)^{2}(x-y)^{2}}+\frac{2(x^{3}+x^{2}y(x-3)+y^{2})\ln(x)}{(1-x)^{3}(x-y)^{3}}-\frac{2y^{2}\ln(y)}{(1-y)(x-y)^{3}}\right),\\
B(x,y)&=\frac{xy}{2(x-y)}\left(\frac{(1-y)(y-3)-2\ln(y)}{(1-y)^{3}}-\frac{(1-x)(x-3)-2\ln(x)}{(1-x)^{3}}\right)-A(y,x),\\
C(x,y)&=\frac{xy}{2}\left(\frac{x+xy+y-3}{(1-x)^{2}(1-y)^{2}}-\frac{2x\ln(x)}{(1-x)^{3}(x-y)} + \frac{2y\ln(y)}{(1-y)^{3}(x-y)}\right).
\end{flalign}
\begin{table}[t]
\begin{tabular}{ |c|c||c|c|c|c|c|c| } 
\hline
&$Q_{EM}$&&$\mathbf{2}_{-1/2}\oplus\mathbf{1}_{-1}$ & $\mathbf{2}_{-1/2}\oplus\mathbf{3}_{-1}$  & $\mathbf{2}_{-3/2}\oplus\mathbf{1}_{-1}$ & $\mathbf{2}_{-3/2}\oplus\mathbf{3}_{-1}$ & $\mathbf{2}_{-1/2}\oplus\;\mathbf{3}_{0}$\\
\hline
$L_{L}^{-}$ & $-1$ & $a_{L_{L}}$& $F(x_{L})$ & $F(x_{L})$ & $F(x_{L}) + B(x_{L},x_{E})$ & $F(x_{L}) + B(x_{L},x_{E})$ & $B(x_{L},x_{E})$\\
\hline
$L_{R}^{-}$  & $-1$ &$a_{L_{R}}$& $F(x_{L})$ & $F(x_{L})$ & $F(x_{L}) + A(x_{L},x_{E})$ & $F(x_{L}) + A(x_{L},x_{E})$ & $A(x_{L},x_{E})$\\
\hline
$L_{L}^{--}$ & $-2$ & $a_{L_{L}}$ & - & - & $F(x_{L})$ & $-F(x_{L}) + 2B(x_{L},x_{E})$  & - \\
\hline
$L_{R}^{--}$ & $-2$ & $a_{L_{R}}$& - & - & $F(x_{L})$ & $-F(x_{L}) + 2A(x_{L},x_{E})$  & - \\
\hline
$E_{L}^{-}$ & $-1$ &$a_{E_{L}}$& $F(x_{E})$ & $F(x_{E})$ & $A(x_{E},x_{L})$ & $A(x_{E},x_{L})$ & $F(x_{E}) + A(x_{E},x_{L})$\\
\hline
$E_{R}^{-}$ & $-1$ &$a_{E_{R}}$& $F(x_{E})$ & $F(x_{E})$ & $B(x_{E},x_{L})$ & $B(x_{E},x_{L})$ & $F(x_{E}) + B(x_{E},x_{L})$\\
\hline
$E_{L}^{--}$ & $-2$ &$a_{E_{L}}$& - & $2F(x_{E})$& - & $2A(x_{E},x_{L})$& - \\
\hline
$E_{R}^{--}$ & $-2$ &$a_{E_{R}}$& - & $2F(x_{E})$& - & $2B(x_{E},x_{L})$& - \\
\hline
$G^{+}$ & $+1$ &$b$& $G(x_{L})$ & $-2G(x_{E}) - G(x_{L})$& $-G(x_{L})$ & $G(x_{L}) - 4C(x_{L},x_{E})$ & $\frac{1}{2}G(x_{L}) + C(x_{L},x_{E})$\\
\hline
\end{tabular}
\caption{Particles with non-zero $Q_{EM}$ in tree models and their contribution to Eq.~\ref{eq:tree_gen}. Entries labeled with - indicate the absence of a particle in a given model. The loop functions needed for each contribution are given in the text.}
\label{table:models_loops}
\end{table}
%

%
\subsection{Loop Models}
Models where $C_{\mu H}$ is generated at one loop without the muon Yukawa coupling comprise UV completions consisting of either two new fermions and one scalar, or two scalars and one fermion. Following~\cite{Capdevilla:2021rwo}, we refer to these cases as FFS- and SSF-models, respectively. We define the couplings and masses in each model via the lagrangians
\begin{figure}[t]
\includegraphics[width=1\linewidth]{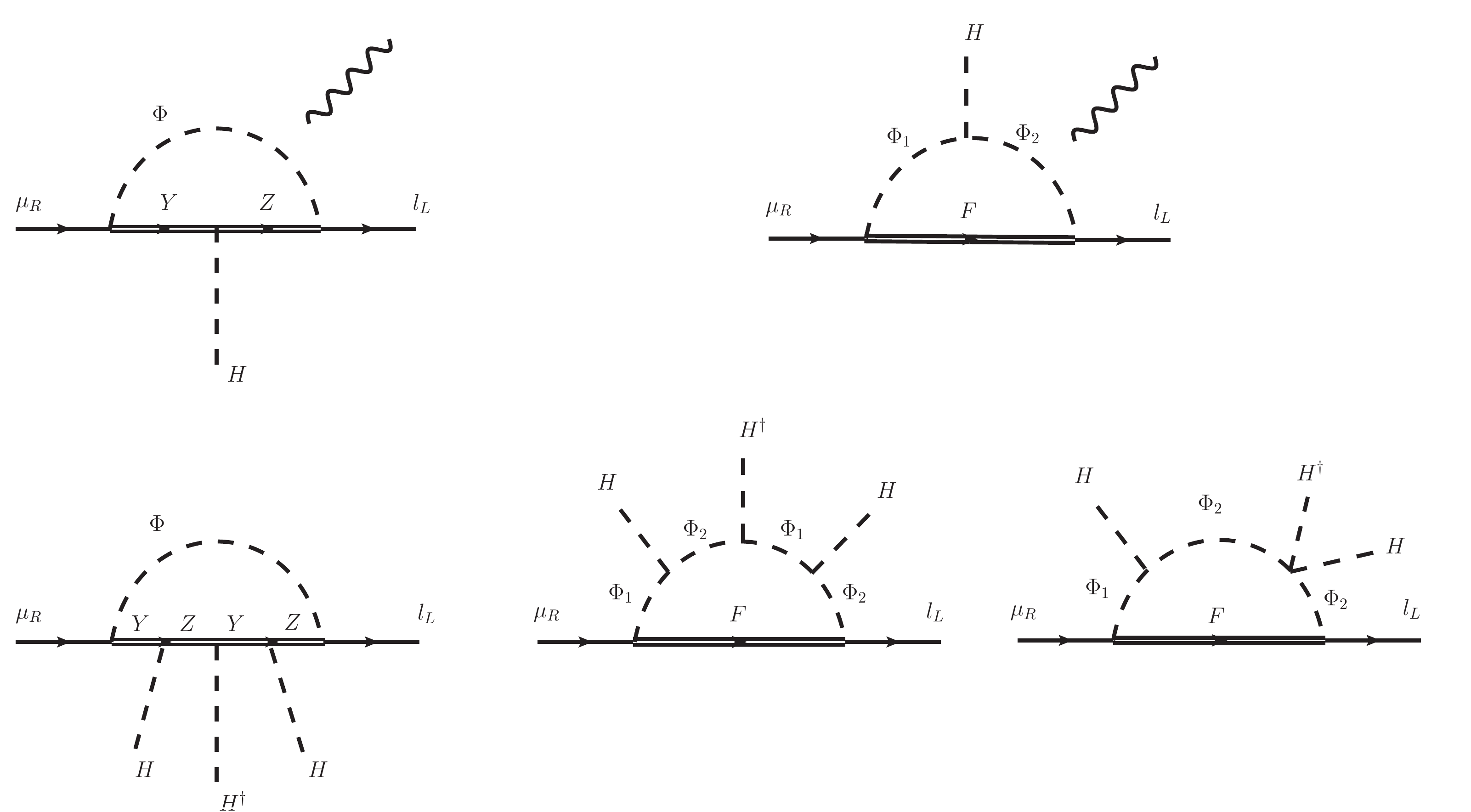}
\caption{Diagrams which generate $C_{\mu H}$ and $C_{\mu \gamma}$ in the FFS- (diagrams with $Y,Z$ fermion lines) and  SSF-type (diagrams with an $F$ fermion line) models considered in~\cite{Calibbi:2018rzv,Crivellin:2018qmi,Crivellin:2021rbq}.}
\label{fig:loop_diags}
\end{figure}
\begin{flalign}\nonumber
-\mathcal{L}_{FFS}\supset& \;\lambda_{Z}\bar{l}_{L}Z_{R}\Phi + \lambda_{Y}\bar{Y}_{L}\mu_{R}\Phi^{\dagger} + \lambda_{YZ}\bar{Z}_{L}Y_{R}H + \lambda^\prime_{YZ}\bar{Z}_{R}Y_{L}H\\
&+ M_{Y}\bar{Y}_{L}Y_{R} + M_{Z}\bar{Z}_{L}Z_{R} + M_{\Phi}^{2}|\Phi|^{2} + h.c.,\label{eq:FFS_lag}\\\nonumber
-\mathcal{L}_{SSF}\supset& \;\lambda_{2}\bar{l}_{L}F_{R}\Phi_{2} + \lambda_{1}\bar{F}_{L}\mu_{R}\Phi_{1}^{\dagger} + A\Phi_{2}^{\dagger}\Phi_{1}H\\
&+M_{F}\bar{F}_{L}F_{R} + M_{\Phi_{1}}^{2}|\Phi_{1}|^{2} + M_{\Phi_{2}}^{2}|\Phi_{2}|^{2} + h.c.,\label{eq:SSF_lag}
\end{flalign}
which for a given model is matched to Eq~\ref{eq:eff_lagrangian}. In the bottom row of Fig.~\ref{fig:loop_diags}, we show the possible diagrams leading to $C_{\mu H}$ where the three Higgs legs are attached either all to the internal fermion line as in the FFS-type models (left), or all on the scalar line as in the SSF-type models (right). For the SSF models we show possible contributions from scalar quartic couplings which may appear depending on details of the model and the scalar potential. Similar diagrams could also possibly appear in FFS models. Charge arrows on internal fermion lines are enforced once specific representations are chosen, as in Fig.~\ref{fig:tree_diags}. It should be noted that in these minimal models there is no tree-level correction to $C_{\mu H}$ even when allowing corrections proportional to the muon Yukawa coupling.

In contrast to the tree models, $C_{\mu\gamma}$ can be constructed starting from the bottom row of Fig.~\ref{fig:loop_diags} by appropriately removing an $H^{\dagger}H$ pair and dressing the resulting diagram with a photon leg.
Following this logic in the FFS scenarios this amounts to removing a factor of $|\lambda_{YZ}|^{2}$ from the diagram for Higgs couplings to fermions as in Eq.~\ref{eq:FFS_lag}, or a factor of $\lambda_{\phi}$ (or $\lambda_{\phi}^{\dagger}$) for scalar quartic couplings. Thus, schematically the expected relation between $C_{\mu\gamma}$ and $C_{\mu H}$ is given by
\begin{equation}
\left(b_{F}|\lambda_{YZ}|^{2}+b_S\lambda_{\phi}\right)C_{\mu \gamma}\simeq\left(\sum_{F}a_{F}Q_{F} + \sum_{S}a_{S}Q_{S}\right)eC_{\mu H},
\label{eq:loop_WC}
\end{equation}
where $a_{S,F}$ are fixed real numbers parameterizing loops in $C_{\mu\gamma}$ with the photon attached to either a scalar or fermion line, respectively. $b_{S,F}$ are analogous numbers parameterizing loops generating $C_{\mu H}$ with or without a scalar quartic coupling, respectively. In the SSF scenarios the same relation is expected with the replacement $\lambda_{YZ}\rightarrow A/M$, where $M$ is identified as a common scale of all new particles. Possible contributions from a scalar quartic coupling, $\lambda_{\phi}$, could be generated from the SM Higgs quartic coupling or otherwise depending on details of the model. In the limit $\lambda_{\phi}\to 0$, we note that although Eq.~\ref{eq:loop_WC} is more complicated than the relation for the tree models, the fact that $C_{\mu H}$ and $C_{\mu\gamma}$ are implicitly determined by $\lambda_{YZ}$ (or $A$) implies that $C_{\mu H}$ and $C_{\mu\gamma}$ are related by a single parameter.

The FFS and SSF-type models have been studied extensively in~\cite{Calibbi:2018rzv,Crivellin:2018qmi,Crivellin:2021rbq} and matching onto SMEFT in the Warsaw basis is provided  in~\cite{Crivellin:2021rbq}. In terms of model building, either case in fact represents an infinite class of models as the couplings alone do not completely determine the quantum numbers of new particles. We follow~\cite{Crivellin:2021rbq} and parameterize the corrections in SSF-type models with the hypercharge of the new fermion, while in the FFS-type models we parameterize corrections in terms of the hypercharge of $\Phi$. However, we recover the results of~\cite{Crivellin:2021rbq} only in the limit $\lambda_{YZ}=\lambda^{\prime}_{YZ}$. We will consider contributions from $\lambda_{YZ}$ and $\lambda^{\prime}_{YZ}$ separately and take the limit $\lambda_{YZ}=\lambda^{\prime}_{YZ}$ as a special case. In each model type, assuming a common scale of new physics (and ignoring contributions from quartic couplings) Eq.~\ref{eq:loop_WC} becomes
\begin{table}[t]
\begin{center}
\begin{tabular}{ |c||c|c|c|c|c| } 
\hline
$\begin{matrix}SU(2)\\FFS\\SSF\end{matrix}$& $\mathcal{Q}_{SSF}$ & $\mathcal{Q}_{FFS}$ & $\mathcal{Q}_{FFS}' $ &$ \mathcal{Q}_{FFS}^{(\lambda_{YZ} = \lambda_{YZ}')}$ & $\xi_{eH}$ \\
\hline
$121$ &$-2(1+2Y_{F})/\xi_{eH}$& $-2(3+4Y_{\Phi})/\xi_{eH}$ & $-2(1+2Y_{\Phi})/3 \xi_{eH}$ & $-2(1+Y_{\Phi})/\xi_{eH}$ & $1$\\
\hline
$212$  & $4Y_{F}/\xi_{eH}$& $2(1+4Y_{\Phi})/\xi_{eH}$ & $4 Y_{\Phi}/3\xi_{eH}$ &$(1+2Y_{\Phi})/\xi_{eH}$ & $-1$\\
\hline
$323$ &$2(1-6Y_{F})/\xi_{eH}$ & $ -2(1+12Y_{\Phi})/\xi_{eH}$& $2(1-6Y_{\Phi}) /3 \xi_{eH}$ &$-2(1+3Y_{\Phi})/\xi_{eH}$& $5$\\
\hline
$232$ & $-4(2+3Y_{F})/\xi_{eH}$ & $-2(11+12Y_{\Phi})/\xi_
{eH}$ & $-4(2+3Y_{\Phi})/3\xi_{eH}$ & $-(7+6Y_{\Phi})/\xi_{eH}$ & $5$\\
\hline
\end{tabular}
\caption{$\mathcal{Q}$ factors defining the relations, Eqs.~\ref{eq:loop_Qs1} and ~\ref{eq:loop_Qs2}, for the FFS- and SSF-type models. In the left-most column, we list the models by representations of new particles under $SU(2)_{L}$. In each case, the leftover hypercharge (as described in the text) is listed as a free parameter in the corresponding row. The right-most column gives the $\xi_{eH}$ factors for a given model, which are fixed by the choice of representations.}
\label{table:loop_models}
\end{center}
\end{table}
\begin{flalign}
e\mathcal{Q}^{(\prime)}_{FFS}C_{\mu H}=4|\lambda^{(\prime)}_{YZ}|^{2}C_{\mu\gamma},\label{eq:loop_Qs1}\\
e\mathcal{Q}_{SSF}C_{\mu H}=4\frac{|A|^{2}}{M^{2}}C_{\mu\gamma},
\label{eq:loop_Qs2}
\end{flalign}
where $\mathcal{Q}^{(\prime)}_{FFS}$,  $\mathcal{Q}_{SSF}$ and $\xi_{eH}$ are determined by the charges of new particles and are defined in Table~\ref{table:loop_models} for the models emphasized in~\cite{Crivellin:2021rbq}.~\footnote{Note that the Wilson coefficients in~\cite{Crivellin:2021rbq} are defined with opposite sign in the Lagrangian compared to our conventions. While this does not affect the definition of the ratio of coefficients in Eqs.~\ref{eq:loop_Qs1} and ~\ref{eq:loop_Qs2}, the sign difference is relevant for matching $C_{\mu\gamma}$ to $\Delta a_{\mu}$.} See the Appendix for more detailed formulas. In the left-most column we list the $SU(2)_{L}$ representations of $Y_{L,R}-Z_{L,R}-\Phi$ particles in Eq.~\ref{eq:FFS_lag} models or the $\Phi_{1}-\Phi_{2}-F_{L,R}$ particles in Eq.~\ref{eq:SSF_lag}. The leftover hypercharge is left as a free parameter for the $\mathcal{Q}$ factor of a given model in the corresponding row. The $\xi_{eH}$ factors are given in the right-most column. For FFS scenarios we give the $\mathcal{Q}$ factors when only $\lambda_{YZ}$ or $\lambda^{\prime}_{YZ}$ is present in the third and fourth columns, respectively. In the fifth column we consider the limit when $\lambda_{YZ} = \lambda_{YZ}'$. In this case, Eq.~\ref{eq:loop_Qs1} has the same form with the replacement $\mathcal{Q}^{(\prime)}_{FFS}\to\mathcal{Q}_{FFS}^{(\lambda_{YZ} = \lambda_{YZ}')}$.

\subsubsection{Bridge Models}
\begin{figure}[t]
\includegraphics[width=1\linewidth]{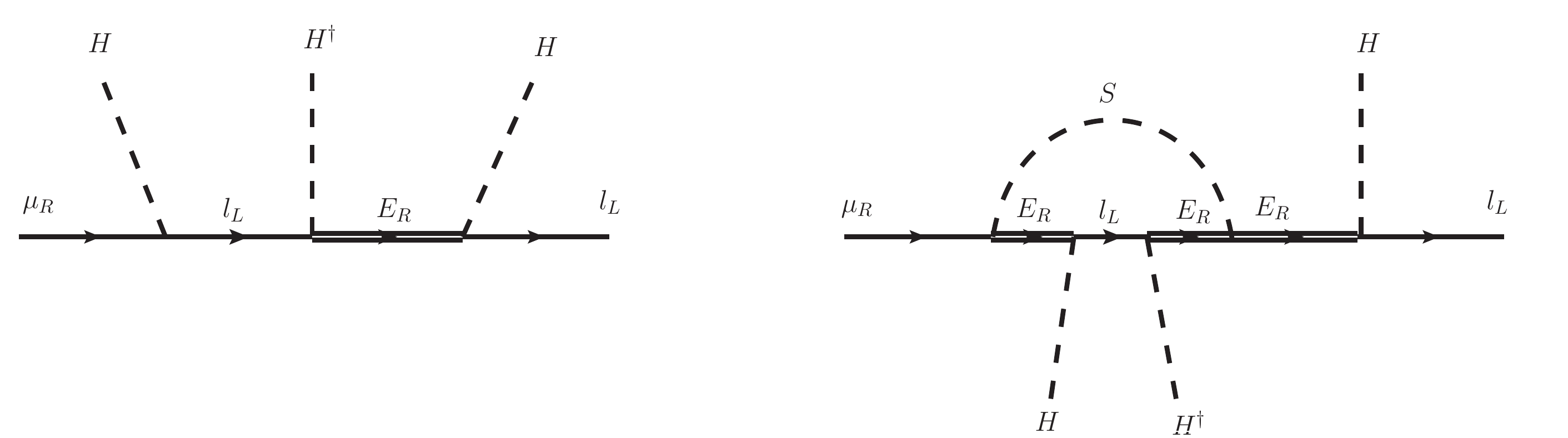}
\caption{Tree and loop diagrams which generate $C_{\mu H}$ in the two-field bridge model introduced in~\cite{Guedes:2022cfy} with a charged fermion singlet, $E_{R}$, and a neutral scalar singlet, $S$.}
\label{fig:bridge_diags}
\end{figure}

The bridge models~\cite{Guedes:2022cfy} represent an interesting class of models sitting somewhere in between the tree and loop models. As an example, we consider the two-field extension of the SM with a new lepton singlet $E\sim(1,1,-1)$ and new scalar $S\sim(1,1,0)$. The relevant couplings of the model are given by
\begin{equation}
-\mathcal{L}\supset y_{\mu}\bar{l}_{L}\mu_{R}H + \lambda_{E}\bar{l}_{L}E_{R}H + \lambda_{S}\bar{E}_{L}\mu_{R}S + \bar{\lambda}\bar{E}_{L}E_{R}S + M_{E}\bar{E}_{L}E_{R} + h.c.
\end{equation}
In this model, $C_{\mu H}$ is generated by a tree-level contribution proportional to the muon Yukawa coupling as well as a loop-level contribution through only new-lepton couplings. The corresponding contributions are generated by the left and right diagrams in Fig.~\ref{fig:bridge_diags}, respectively, and give
\begin{equation}
C_{\mu H} \simeq \frac{y_{\mu}|\lambda_{E}|^{2}}{M_{E}^{2}} - \frac{\lambda_S\lambda_{E}\bar{\lambda}}{16\pi^{2}}\frac{|\lambda_{E}|^{2}}{M_{E}^{2}}D(x),
\label{eq:CmuH_bridge}
\end{equation}
where
\begin{equation}
D(x)=\frac{x^{2}-x-x\ln(x)}{(1-x)^{2}},
\end{equation}
 and $x=M_{E}^{2}/M_{S}^{2}$.~\footnote{When $\lambda_{S}\lambda_{E}\bar{\lambda}>0$, there is also the possibility that $C_{\mu H}=0$. Although, exact cancellation would require some fine tuning in the model.} Taking $M_{E}\sim M_{S}$ we have that 
\begin{equation}
\frac{C^{(loop)}_{\mu H}}{C^{(tree)}_{\mu H}}\sim \frac{\lambda_S\lambda_{E}\bar{\lambda}v}{32\pi^{2}m_{\mu}},
\end{equation}
which is maximal when the bound from the muon coupling to the $Z$-boson, $|\lambda_{E}|v/M_{E}\leq 0.03$, is saturated. When this occurs, the loop contribution can dominate for $\lambda_{S},\bar{\lambda}\simeq 1.0 - 0.5$
and a scale of new physics at $M_{E}\sim M_{S} = 1-10$ TeV. Thus, in the two-field extensions of the bridge models it is easy for the dominant contribution to $C_{\mu H}$ to be that generated by the loop in Fig.~\ref{fig:bridge_diags} (right), for moderate-size new couplings up to the size allowed by perturbativity limits at the scale of new physics. 

In this model, $C_{\mu\gamma}$ is constructed from the right diagram by removing the Higgs vertices in the loop and dressing the resulting diagram with an external photon. Further, we have that (c.f. Eq. 4.7 of~\cite{Guedes:2022cfy})
\begin{equation}
C_{\mu\gamma} \simeq e\frac{\lambda_{S}\lambda_{E}\bar{\lambda}}{32\pi^{2}M_{E}^{2}}F(x),
\end{equation}
where $F(x)$ is defined in Eq.~\ref{eq:Ffunc}. In the region of parameters where the loop contribution to $C_{\mu H}$ dominates, we have
\begin{equation}
3|\lambda_{E}|^{2}C_{\mu \gamma}\simeq e C_{\mu H}.
\label{eq:bridge_example}
\end{equation}

Beyond this example, in~\cite{Guedes:2022cfy} it was found that there are five additional such models which generate a mass-enhanced correction to $\Delta a_{\mu}$ via the same topology.~\footnote{There are eight possible models in total. However, two possibilities result in $C_{\mu\gamma}=0$ regardless of the hierarchy of new masses involved. It would be interesting to investigate if these cases share a similar type of internal symmetries which lead to a "magic zero" as in~\cite{Craig:2021ksw,DelleRose:2022ygn}.} In each case, the same arguments correlating $C_{\mu H}$ would apply albeit with a slightly different numerical factor in Eq~\ref{eq:bridge_example}. Thus, in the six examples we may write a generic expression as
\begin{equation}
e \mathcal{Q}C_{\mu H}\simeq|\lambda_{F\mu}|^{2}C_{\mu \gamma},
\label{eq:bridge_equation}
\end{equation}
\begin{table}[t]
\begin{center}
\begin{tabular}{ |c||c| } 
\hline
$SU(2)\times U(1)_{Y}$& $\mathcal{Q}$ \\
\hline
$\mathbf{1}_{-1}\oplus\mathbf{1}_{0}$ & $-1/3$ \\
\hline
$\mathbf{1}_{-1}\oplus\mathbf{1}_{-2}$  & $2/3$\\
\hline
$\mathbf{2}_{-1/2}\oplus\mathbf{1}_{0}$ & $-1/3$ \\
\hline
$\mathbf{2}_{-1/2}\oplus\mathbf{3}_{0}$ & - \\
\hline
$\mathbf{2}_{-1/2}\oplus\;\mathbf{3}_{-1}$ & $3/4$ \\
\hline
$\mathbf{3}_{-1}\oplus\;\mathbf{3}_{0}$ & $-1/2$ \\
\hline
\end{tabular}
\caption{Quantum numbers of $F\oplus S$ under $SU(2)\times U(1)_{Y}$ and corresponding $\mathcal{Q}$-factor relating $C_{\mu\gamma}$ and $C_{\mu H}$ in the two-field bridge models~\cite{Kannike:2011ng}.}
\label{table:bridge_models}
\end{center}
\end{table}
where, in a given model, $\lambda_{F\mu}$ should be understood as the coupling of the Higgs which mixes the heavy fermion with either the left- or right-handed muon. In Table~\ref{table:bridge_models}, we provide a list of the possible two-field bridge models which generate a mass-enhanced correction to $\Delta a_{\mu}$. In the left column, we give the quantum numbers of the new fermion and scalar ($F\oplus S$) under $SU(2)\times U(1)_{Y}$ and in the right column we list the corresponding $\mathcal{Q}$-factor in each case appearing in Eq.~\ref{eq:bridge_equation}. For the model $\mathbf{2}_{-1/2}\oplus\mathbf{3}_{0}$, $C_{\mu\gamma}$ happens to vanish in the limit $M_{\mathbf{2}_{-1/2}}=M_{\mathbf{3}_{0}}$. However, for a general spectrum of new particles the relation to $C_{\mu H}$ is given by
\begin{equation}
4|\lambda_{F\mu}|^{2}C_{\mu\gamma}=e\frac{H(x)}{D(x)}C_{\mu H},
\end{equation}
where
\begin{equation}
H(x)=-\frac{x^{3} + 4x^{2} - 5x - 2x(2x + 1)\ln(x)}{(1-x)^3},
\end{equation}
and $\sqrt{x}=M_{\mathbf{2}_{-1/2}}/M_{\mathbf{3}_{0}}$.

We see that in the 2-field models discussed in~\cite{Guedes:2022cfy}, $C_{\mu\gamma}$ is correlated with the loop contribution to $C_{\mu H}$ through only a single parameter, Eq.~\ref{eq:bridge_equation}, similarly as in the loop models. Further, the loop contribution to $C_{\mu H}$ can dominate in regions of parameter space with the largest couplings and highest mass scales. We note that the 3-field bridge models discussed in~\cite{Guedes:2022cfy} do not exhibit this correlation and are the only class of mass-enhanced models suggested as solutions for g-2 which evade this feature at the level we have outlined here. However, we do find that in these scenarios the correlation between $C_{\mu H}$ and $C_{\mu\gamma}$ can appear at the two-loop level by connecting the diagrams generating $C^{(1)}_{Hl}H^{\dagger}i\overset{\leftrightarrow}{D_{\mu}}H\left(\bar{l}_{L}\gamma^{\mu}l_{L}\right)$, $C^{(3)}_{Hl}H^{\dagger}i\overset{\leftrightarrow}{D_{\mu}^{I}}H\left(\bar{l}_{L}\tau^{I}\gamma^{\mu}l_{L}\right)$, or $C_{He}H^{\dagger}i\overset{\leftrightarrow}{D_{\mu}}H\left(\bar{\mu}_{R}\gamma^{\mu}\mu_{R}\right)$ to that of $C_{\mu\gamma}$ on an external fermion line. We do not explore this in this paper as these corrections are not expected to compete with the tree level correction, as in Eq.~\ref{eq:CmuH_bridge}.
\section{The ellipse of dipole moments}
\label{sec:ellipse}
In the tree and loop models discussed above, we have argued that the dominant contribution to the dipole operator is parametrically related to the correction to the muon-Higgs coupling by
\begin{equation}
C_{\mu H}\simeq \frac{k}{e} C_{\mu\gamma},
\label{eq:WC_relation}
\end{equation}
for some numerical factor $k$ which depends on the details of the model. Further, this relation applies to the 2-field bridge models in the moderate to extreme regions of parameter space. In the effective theory, regardless if the operators are generated in tree or loop models, this relation has important consequences for predictions of the muon dipole moments and $h\to\mu^{+}\mu^{-}$ branching ratio. Assuming the generic relation, Eq.~\ref{eq:WC_relation}, in Eq~\ref{eq:R_mu_def} and using Eqs~\ref{eq:mdipole} and \ref{eq:edipole} we find that the modification of the Higgs decay to muons defines an ellipse with respect to the electric and magnetic dipole moments

%
%
\begin{flalign}
R_{h\to\mu^{+}\mu^{-}}=\left(\frac{\Delta a_{\mu}}{2\omega} - c_{\phi_k}\right)^{2} + \left(\frac{m_{\mu}d_{\mu}}{e\omega} - s_{\phi_k}\right)^{2},
\label{eq:ellipse}
\end{flalign}
where $c_{\phi_k}=\cos \phi_k$, $s_{\phi_k}=\sin \phi_k$ and $\omega = m_{\mu}^{2}/|k|v^{2}$. Thus, in models where $C_{\mu\gamma}$ is generated by mass-enhanced corrections future measurements of $R_{h\to\mu^{+}\mu^{-}}$ and $\Delta a_{\mu}$ are directly connected to the determination of the electric dipole moment, $d_{\mu}$.  

\begin{figure}[t]
\includegraphics[scale=0.5]{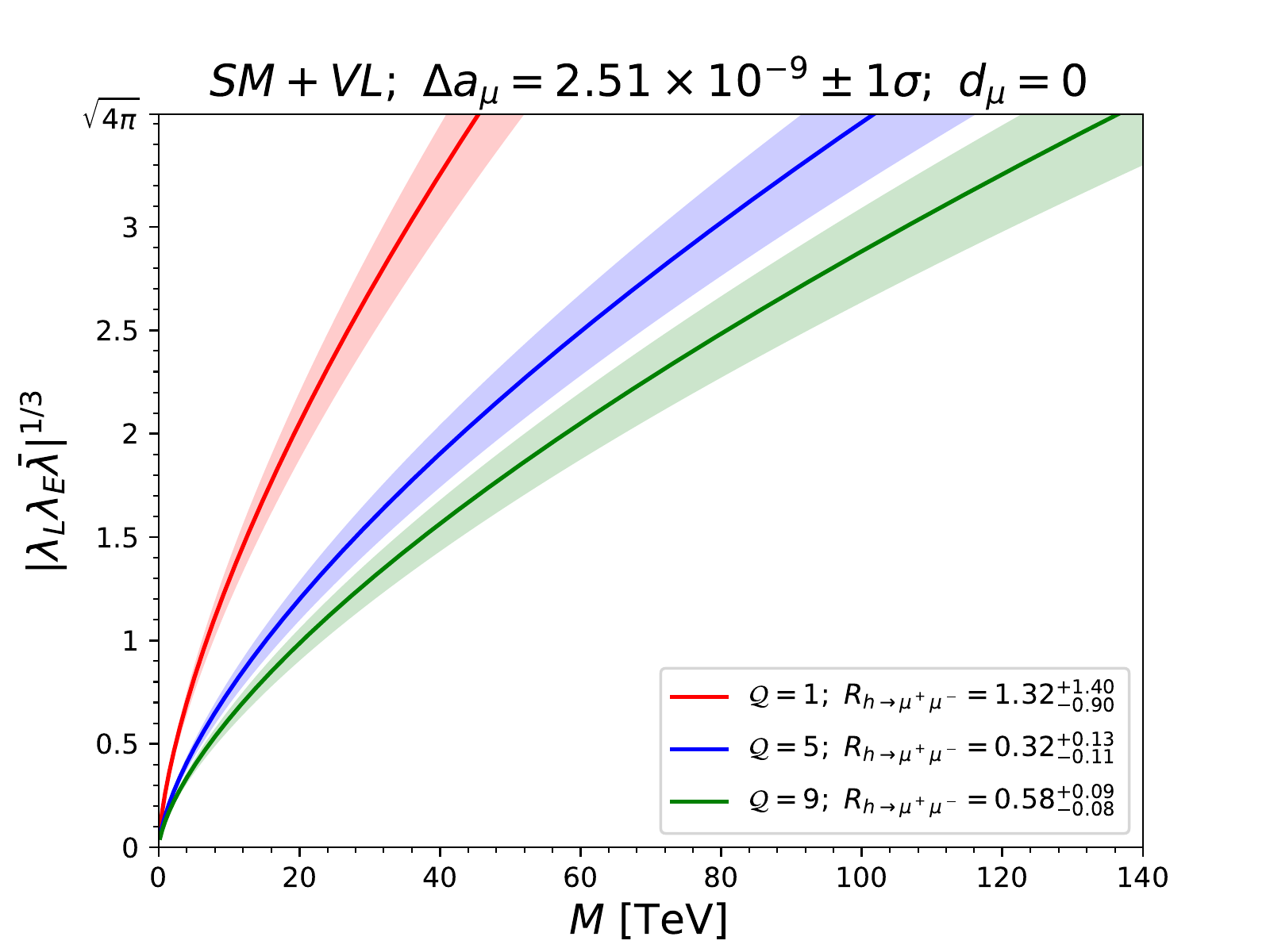}
\includegraphics[scale=0.5]{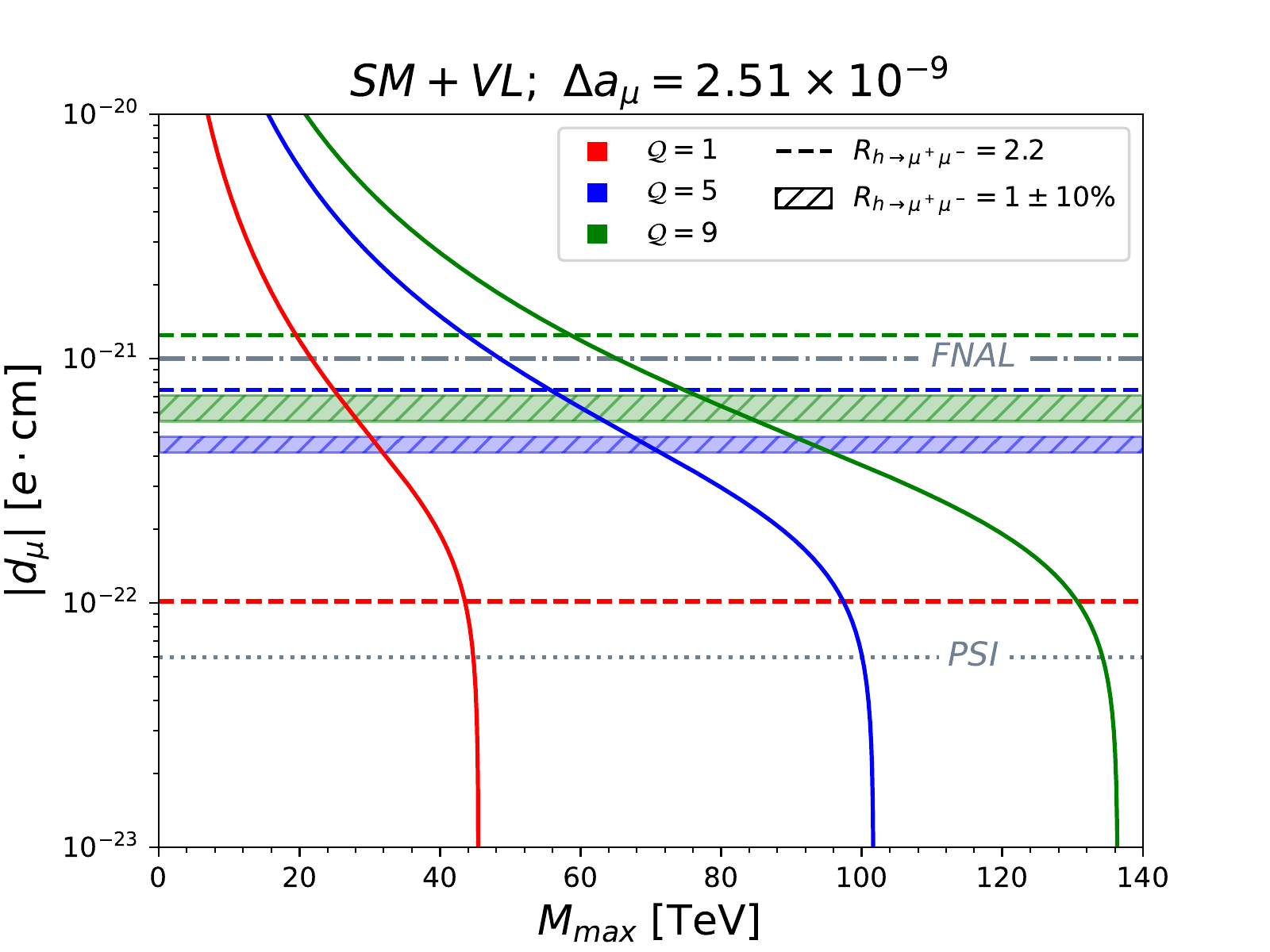}
\caption{Left: Common scale of new physics required in tree models with $\mathcal{Q}=1, 5,9$ needed to achieve $\Delta a_{\mu}\pm 1\sigma$ for a given overall size of model couplings, assuming $d_{\mu}=0$. Right: Values of $|d_{\mu}|~[e\cdot cm]$ with respect to the maximum mass scale allowed to give the central value of $\Delta a_{\mu}$ in tree models. Dashed lines show the value of $d_{\mu}$ as given by the $k$ equation when $R_{h\to \mu^{+}\mu^{-}}= 2.2$ for a given model. Hatched regions show the corresponding range for $R_{h\to \mu^{+}\mu^{-}}= 1\pm 10\%$. Projected sensitivity to $|d_{\mu}|$ from Fermilab and PSI are shown with dash-dotted and dotted lines, respectively.}
\label{fig:tree_scales}
\end{figure}
Concretely, for tree models, assuming a single scale of new physics, we have for Eq.~\ref{eq:WC_relation} 
\begin{equation}
k=-\frac{1}{\left[\sum_{j=L,R}\sum_{i=L,E}Q_{i_{j}}F(x_{i}) + Q_{G^{+}}G(x_{L})\right]}=\frac{64\pi^{2}}{\mathcal{Q}},
\label{eq:tree_X_SM}
\end{equation}
where $\mathcal{Q}$ is a numerical factor for a given model as in Table~\ref{table:models}. In Fig.~\ref{fig:tree_scales}, we show mass scale of new physics, assumed to be common among all new particles, required in the tree models with $\mathcal{Q}=1, 5,9$ to obtain $\Delta a_{\mu}\pm 1\sigma$ for an overall size of couplings, defined as $(|\lambda_{L}\lambda_{E}\bar{\lambda}|)^{1/3}$. In the left panel, $d_{\mu}=0$. In the legend, we also quote the predicted range of $R_{h\to \mu^{+}\mu^{-}}$ for a given $\mathcal{Q}$. In each case, we have checked that EW precision constraints are satisfied. For $\mathcal{Q}=1$, mass scales up to $\sim 45$ TeV can explain $\Delta a_{\mu}\pm 1\sigma$ before the overall size of couplings reaches non-perturbative values. For $\mathcal{Q}=9$, mass scales up to slightly above 135 TeV are possible under the same restriction.

In the right panel of Fig.~\ref{fig:tree_scales}, we show the maximum mass scale allowed in tree models which explain the central value of $\Delta a_{\mu}$ as a function of the predicted value of $d_{\mu}$. For each model, a given value for $d_{\mu}$ automatically fixes $R_{h\to\mu^{+}\mu^{-}}$. The dashed lines show the values of $d_{\mu}$ corresponding to $R_{h\to \mu^{+}\mu^{-}}= 2.2$ in a given model. Hatched regions show the corresponding range for $R_{h\to \mu^{+}\mu^{-}}= 1\pm 10\%$. Note that for $\mathcal{Q}=1$, the region where $R_{h\to \mu^{+}\mu^{-}}= 1\pm 10\%$ never occurs. Even for $d_{\mu} = 0$, the smallest value of $R_{h\to \mu^{+}\mu^{-}}$ is 1.32, assuming the central value of $\Delta a_{\mu}$. Interestingly, this implies that if future measurements of Higgs decays reach $R_{h\to \mu^{+}\mu^{-}}= 1\pm 10\%$, a measurement at FNAL or PSI of $|d_{\mu}|\neq 0$ up to their projected limits will rule out models with $\mathcal{Q}=1$, assuming the central value of $\Delta a_{\mu}$. However, values as low as $R_{h\to \mu^{+}\mu^{-}}=0.42$ are possible for $\Delta a_{\mu}-1\sigma$. Projected sensitivity to $|d_{\mu}|$ from Fermilab and PSI are shown with dash-dotted and dotted lines, respectively. Note that while models with $\mathcal{Q}=5,9$ would not be ruled out by FNAL, their range of validity as evaluated by the $k$ equation would be severely restricted, requiring $|d_{\mu}|\simeq 4.5\times 10^{-22}\; [e\cdot cm]$ and $|d_{\mu}|\simeq 6\times 10^{-22}\; [e\cdot cm]$, respectively. Again assuming $R_{h\to \mu^{+}\mu^{-}}= 1\pm 10\%$, if $|d_{\mu}|$ is \textit{not} seen at PSI, models with $\mathcal{Q}=5,9$ would be completely excluded.

%

For the loop models, we discussed in previous sections minimal UV completions involving either two new fermions and one scalar, or two new scalars and one fermion. For the FFS-type models, assuming a single scale of new physics, we have for Eq.~\ref{eq:WC_relation} 
\begin{equation}
k=\frac{\left(b_{F}|\lambda_{YZ}|^{2}+b_S\lambda_{\phi}\right)}{\left(\sum_{F}a_{F}Q_{F} + \sum_{S}a_{S}Q_{S}\right)},
\label{eq:loop_k}
\end{equation}
and similarly in the SSF models with the replacement $\lambda_{YZ}\to A/M$, as discussed above. Ignoring, for the moment, contributions from quartic couplings we can identify the appropriate $k$ factors as 
\begin{flalign}
k_{FFS}=&\frac{4}{\mathcal{Q}^{(\prime)}_{FFS}}|\lambda^{(\prime)}_{YZ}|^{2},\\
k_{SSF}=&\frac{4}{\mathcal{Q}_{SSF}}\frac{|A|^{2}}{M^{2}},
\end{flalign}
where the $\mathcal{Q}$'s and $\xi_{eH}$ factors are given in Table~\ref{table:loop_models}. The relation in the FFS models when $\lambda_{YZ} = \lambda_{YZ}'$ as discussed above would have the same form. We discuss corrections due to quartic couplings and other generalizations in the next section.

Finally, in the 2-field extensions of the bridge models we have

\begin{equation}
k=\frac{|\lambda_{F\mu}|^{2}}{\mathcal{Q}},
\end{equation}
where the $\mathcal{Q}$ values are given in Table~\ref{table:bridge_models}. We reiterate that this relation applies in regions of parameters where the loop contribution to $C_{\mu H}$ is dominant. This also corresponds to the region of parameters where a given model gives the largest contribution to $\Delta a_{\mu}$.
\begin{figure}[t]
\includegraphics[scale=0.65]{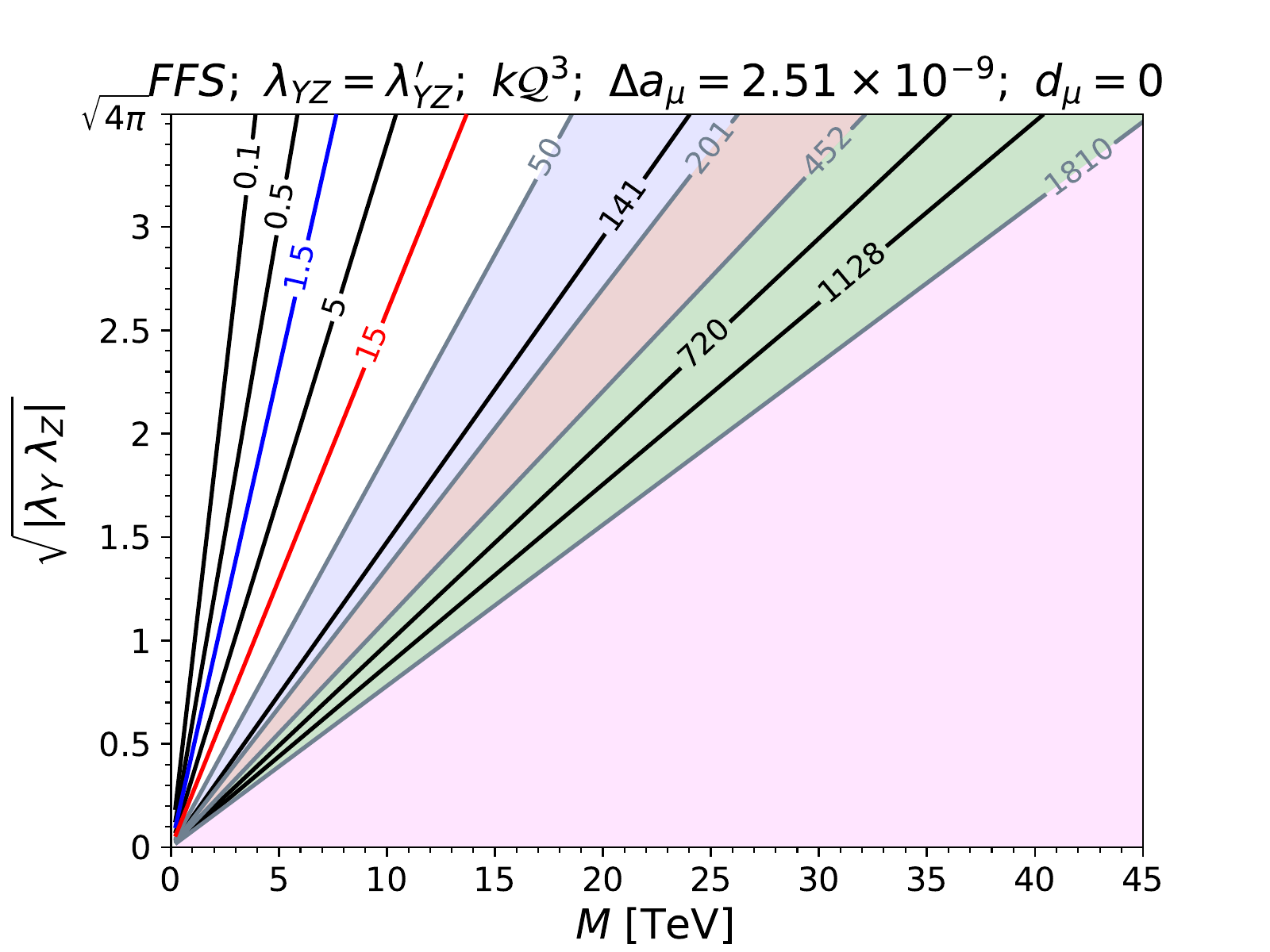}
\caption{Contours of $k\mathcal{Q}^{3}$ in the FFS models with $SU(2)$ doublets and $\lambda_{YZ} = \lambda_{YZ}'$, as in the first two rows and fifth column of Table~\ref{table:loop_models}, defined in the plane of the scale of new physics needed to obtain the central value of $\Delta a_{\mu}$ with respect to the overall size of model couplings. The blue shaded region is excluded by the perturbativity limit $|\lambda_{YZ}|\leq \sqrt{4\pi}$ for $\mathcal{Q}=1$. The red, green, and purple regions show the corresponding exclusion regions for $\mathcal{Q}=2$, $\mathcal{Q}=3$ and $\mathcal{Q}=6$, respectively. The red and blue contours are highlighted to show where $R_{h\to \mu^{+}\mu^{-}}=0.9$ and $0.99$ for $\mathcal{Q}=1$, respectively.} 
\label{fig:scales1}
\end{figure}

To demonstrate the correlation between $\Delta a_{\mu}$, $R_{h\to\mu^{+}\mu^{-}}$, and $d_{\mu}$ in FFS models we may write the dipole moments with respect to $k$ by
\begin{flalign}
\Delta a_{\mu}=&\frac{1}{96\pi^{2}}\left(\frac{m_{\mu}v}{M^{2}}\right)\left(\frac{k\mathcal{Q}_{FFS}^{3}\xi_{eH}^{2}}{4}\right)^{1/2}\textrm{Re}\left[\lambda_{Y}\lambda_{Z}e^{i\phi_{YZ}}\right],\label{eq:amu_k_loop}\\
|d_{\mu}|=&\frac{e}{192\pi^{2}}\left(\frac{v}{M^{2}}\right)\left(\frac{k\mathcal{Q}_{FFS}^{3}\xi_{eH}^{2}}{4}\right)^{1/2}\textrm{Im}\left[\lambda_{Y}\lambda_{Z}e^{i\phi_{YZ}}\right],\label{eq:dmu_k_loop}
\end{flalign}
where we have assumed a common scale of new physics and the $\xi_{eH}$ factors are defined in Table~\ref{table:loop_models}. For SSF models, the formulas are similar except with an overall sign in the analogous relation for $\Delta a_{\mu}$. We reiterate that the combined factor $Q^{3}\xi_{eH}^{2}$ is dependent only on the choice of representions of new particles and is fixed for a given model.
In Fig.~\ref{fig:scales1}, we show contours of $k\mathcal{Q}^{3}$ for the FFS models with $\lambda_{YZ} = \lambda_{YZ}'$ in the plane of the common scale of new physics for a given overall size of couplings, this time defined as $\sqrt{|\lambda_{Y}\lambda_{Z}|}$ where we have eliminated the free parameter $\lambda_{YZ}$ in favor of $k\mathcal{Q}^{3}$. We have used the central value of $\Delta a_{\mu}$ and assumed $d_{\mu}=0$. The contours are drawn for models with $SU(2)$ doublets, as in the first two rows of Table~\ref{table:loop_models}. For models with triplets as in the bottom two rows, the contour labels should be multiplied  by a factor of 25, resulting from the modified factors of $\mathcal{Q}^{3}\xi_{eH}^{2}$ in Eqs.~\ref{eq:amu_k_loop} and~\ref{eq:dmu_k_loop}. A larger scale of new physics requires a larger size couplings and hence larger $k\mathcal{Q}^{3}$. For example, we see that a new physics scale of $\sim 10$ TeV would require at most $\mathcal{O}(1)$ couplings. In an FFS model, increasing $k\mathcal{Q}^{3}$ is achieved by adjusting $\lambda_{YZ}$ while respecting the limit on perturbativity at the scale of new physics, $|\lambda_{YZ}|\leq \sqrt{4\pi}$. For $\mathcal{Q}=1$, this coupling becomes nonperturbative in the shaded blue region. Similarly the red, green, and purple shaded regions show the analogous exclusion region for models with $\mathcal{Q}=2$, $\mathcal{Q}=3$ and $\mathcal{Q}=6$, respectively. For models with $\lambda_{YZ}=0$ ($\lambda^{\prime}_{YZ}=0$)  the contours should be rescaled by $9/4~(1/4)$. The corresponding figure for SSF models looks similar with all contour labels divided by a factor of $4$ resulting from an additional factor of $1/4$ in the square root factors of Eqs.~\ref{eq:amu_k_loop} and~\ref{eq:dmu_k_loop}. We note that $k\mathcal{Q}^{3}\propto |A|^{2}/M^{2}$ in SSF models and there is no analogous perturbativity limit on $A$. 

\begin{figure}[t]
\includegraphics[scale=0.5]{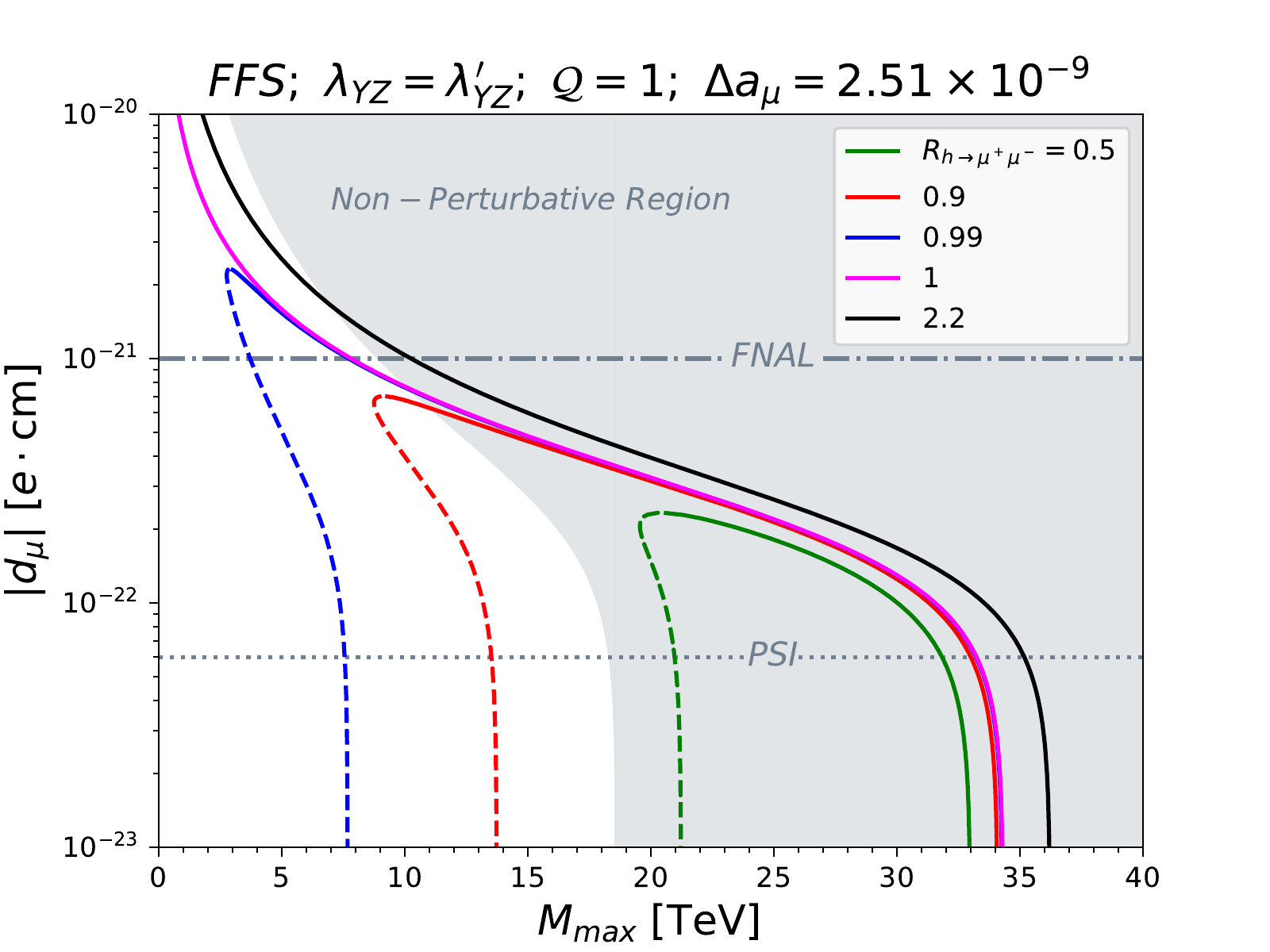}
\includegraphics[scale=0.5]{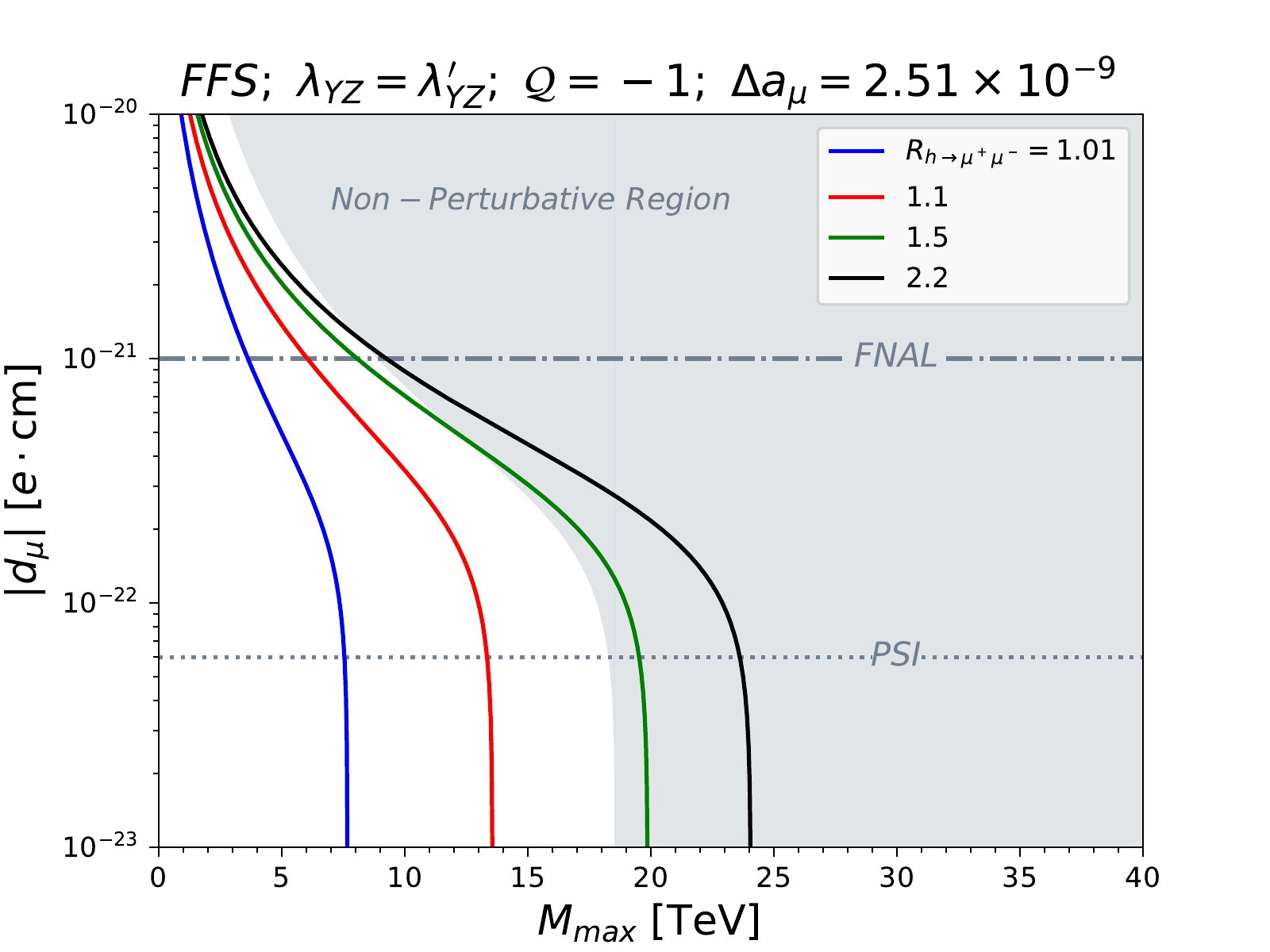}
\caption{Contours of $R_{h\to\mu^{+}\mu^{-}}$with respect to $|d_{\mu}|$ and the maximum mass scale allowed by the correlation with $\Delta a_{\mu}$ given by the $k$ equation in the FFS models with $\lambda_{YZ} = \lambda_{YZ}'$ and $\mathcal{Q}=+1$ (left) and $\mathcal{Q}=-1$ (right). Projected sensitivity to $|d_{\mu}|$ from Fermilab and PSI are shown with dash-dotted and dotted lines, respectively. The gray shaded region shows where all model couplings would be nonpertubrative.}
\label{fig:scales2}
\end{figure}

The red and blue contours in Fig.~\ref{fig:scales1} are highlighted to show where $R_{h\to \mu^{+}\mu^{-}}=0.9$ and $0.99$, respectively, when $\mathcal{Q}=1$. Note that these highlighted contours show that increasing precision in $R_{h\to \mu^{+}\mu^{-}}$ provides an upper limit on the scale of new physics which is stronger than that based on pertubativity, and that this upper limit decreases as $R_{h\to \mu^{+}\mu^{-}}\to1$ . Further, recall that $d_{\mu}\neq 0$ can only increase $R_{h\to \mu^{+}\mu^{-}}$ in the models we have discussed. Thus, for $d_{\mu}\neq 0$ it is expected that a given precision on $R_{h\to \mu^{+}\mu^{-}}$ provides an upper limit on the scale of new physics which is strictly less than the corresponding maximum scale allowed as shown in Fig.~\ref{fig:scales1}. We show the maximum scale allowed defined in this way in Fig.~\ref{fig:scales2} with respect to $|d_{\mu}|$. The left (right) panel shows contours for the FFS models with $\mathcal{Q}=+1(-1)$. Note that the sign of $\mathcal{Q}$ is always the same as $k$. Thus, for $\mathcal{Q}=+1$ Eq.~\ref{eq:ellipse} allows two positive solutions for $k$ when $R_{h\to\mu^{+}\mu^{-}}<1$ whose contours are denoted by the solid and dashed lines. In contrast, Eq.~\ref{eq:ellipse} only admits one positive solution when $R_{h\to\mu^{+}\mu^{-}}>1$ for $\mathcal{Q}=+1$. For $\mathcal{Q}=-1$, there is only one solution for $k$ and we have that $R_{h\to\mu^{+}\mu^{-}}>1$. We also show the projected upper limits on $|d_{\mu}|$ from the experiment at Fermilab (dash-dotted) and PSI (dotted). The gray shaded regions show where all the model couplings reach the limit of perturbativity at the scale $M_{max}$.

Focusing first on the left panel, $\mathcal{Q}=+1$, we see that for a given value of $|d_{\mu}|$, the current limit $R_{h\to\mu^{+}\mu^{-}}<2.2$ defines an upper limit on the scale of new physics up to $\sim 8$ TeV before the model couplings become nonperturbative. Slightly higher scales are allowed depending on the future precision of Higgs decay measurements. In particular, allowing for $R_{h\to\mu^{+}\mu^{-}}$ within 10\% of the SM value, we see that the upper limit on the scale of new physics does not exceed $\sim 14$ TeV for any value of $|d_{\mu}|$. For $\mathcal{Q}=-1$ we see that $M_{max}$ is strictly decreasing as $R_{h\to\mu^{+}\mu^{-}}\to 1$. Similarly, for $R_{h\to\mu^{+}\mu^{-}}$within 10\% of the SM value, the upper limit on the scale of new physics is $\sim 14$ TeV.
\begin{figure}[t]
\includegraphics[scale=0.5]{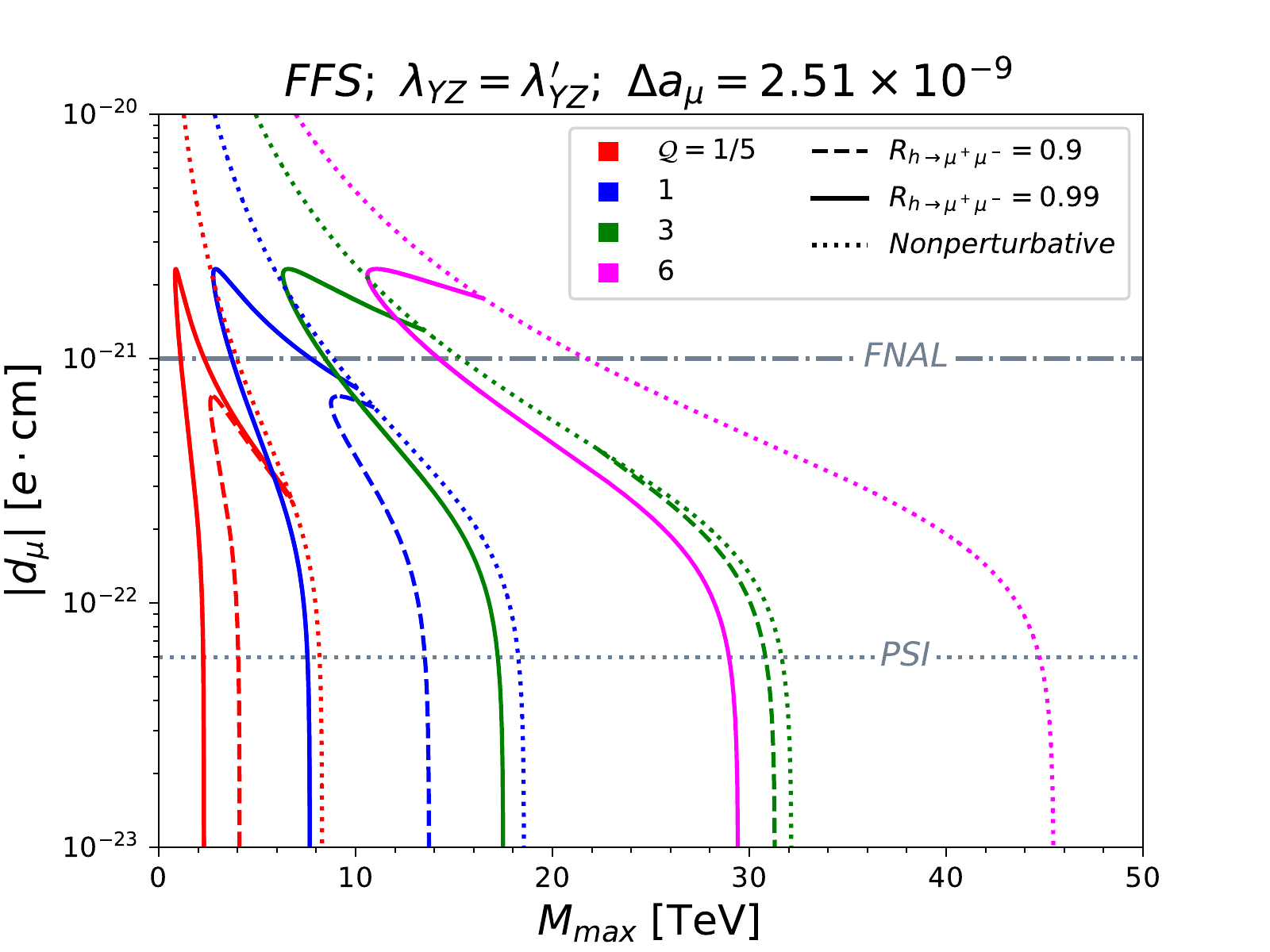}
\includegraphics[scale=0.5]{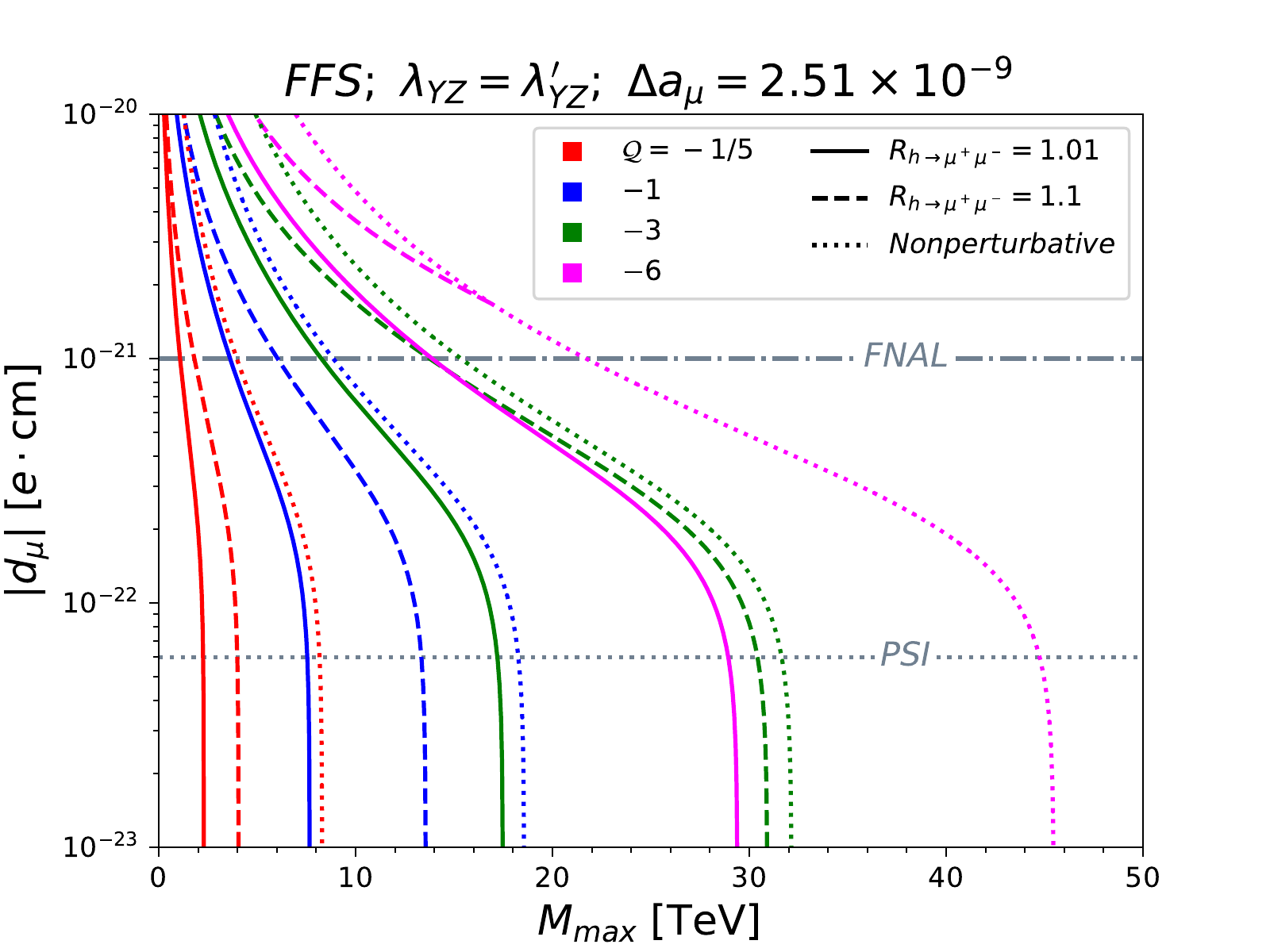}\\
\includegraphics[scale=0.5]{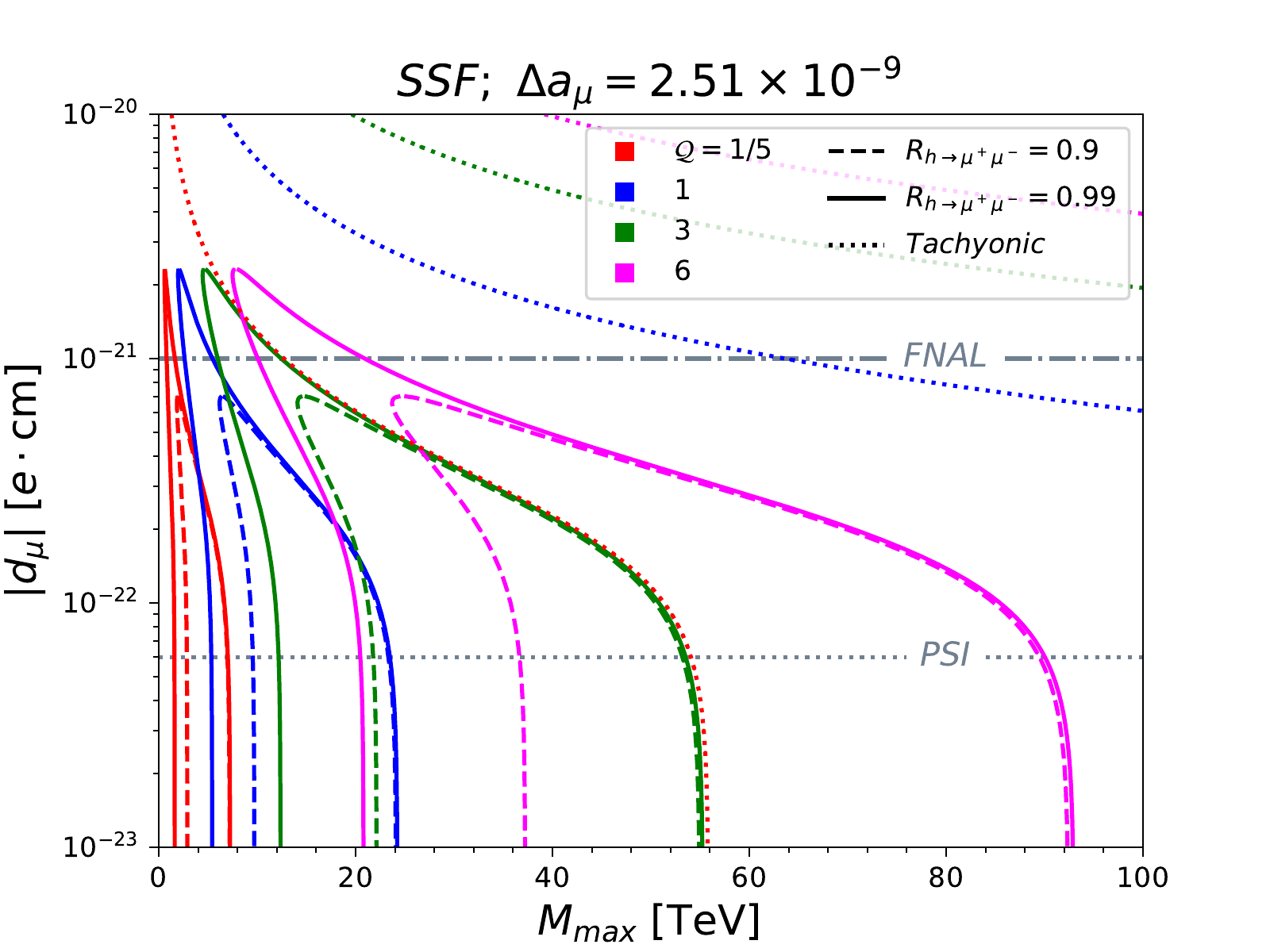}
\includegraphics[scale=0.5]{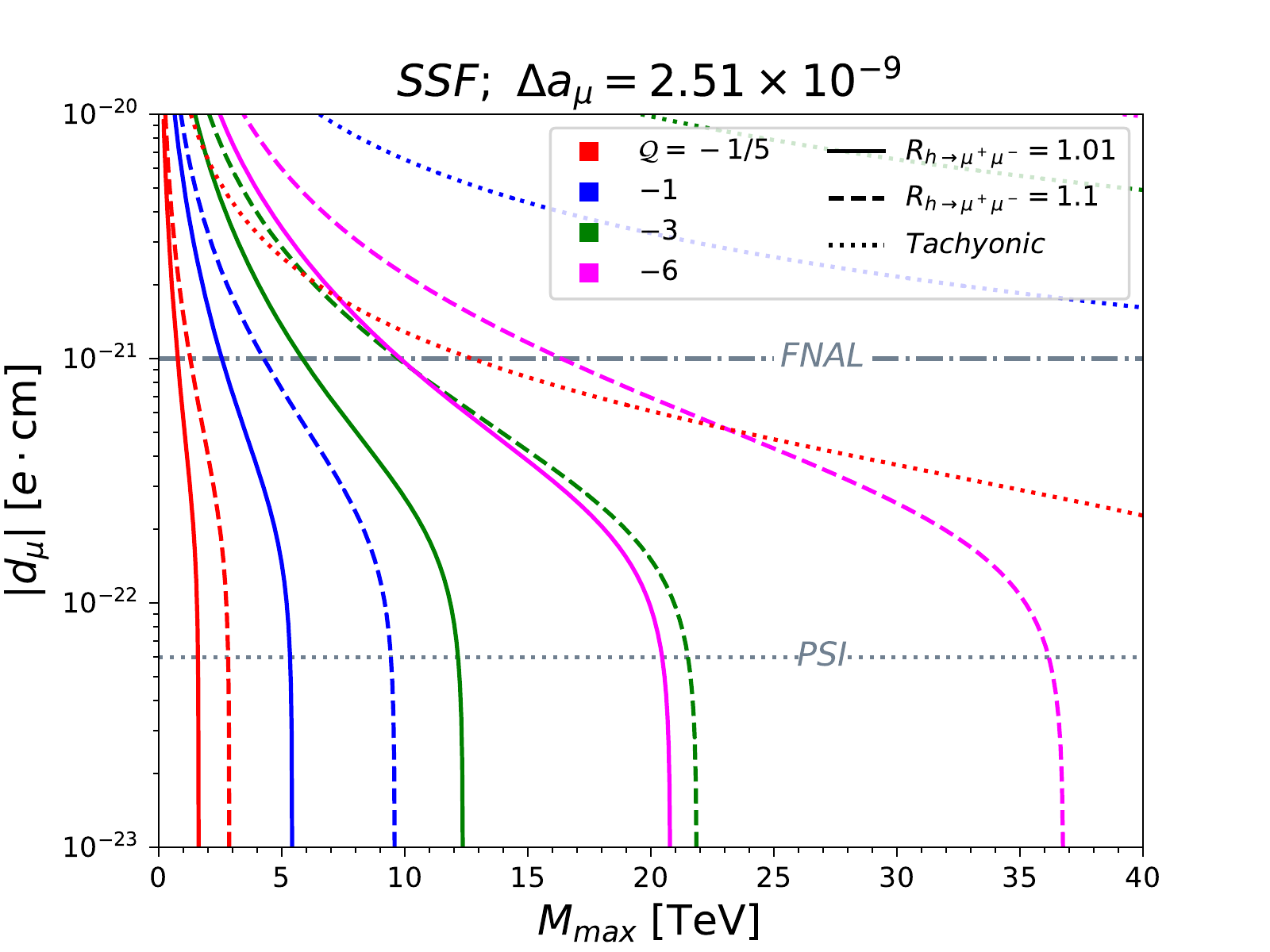}
\caption{Contours of $R_{h\to\mu^{+}\mu^{-}}$with respect to $|d_{\mu}|$ and the maximum mass scale as in Fig.~\ref{fig:scales2} for different values of $\mathcal{Q}$. The top row shows contours in the FFS models with $\lambda_{YZ}=\lambda_{YZ}^{\prime}$ for $\mathcal{Q}>0$ (left) and $\mathcal{Q}<0$ (right). The bottom row shows the corresponding contours in the SSF models.}
\label{fig:scales3}
\end{figure}

In Fig.~\ref{fig:scales3}, we show similar contours as in Fig.~\ref{fig:scales2} for various values of $\mathcal{Q}$. In the top row, we show the upper limits on the scale of new physics in the FFS models for $\mathcal{Q}>0$ (left) and $\mathcal{Q}<0$ (right). In each case we show the upper limit on the scale of new physics for $R_{h\to \mu^{+}\mu^{-}}=0.9$ (dashed) and $R_{h\to \mu^{+}\mu^{-}}=0.99$ (solid). The dotted lines show the regions of nonpertubativity for a given model.  We see that, within the parameter space allowed by perturbativity, increasing precision in $R_{h\to \mu^{+}\mu^{-}}$ can have an impact on the scale of new physics allowed for a given prediction of $|d_{\mu}|$. For instance, for models with $\mathcal{Q}=+1$ which will satisfy the PSI bound on $|d_{\mu}|$, $R_{h\to \mu^{+}\mu^{-}}=0.9\to0.99$ decreases the maximum scale of new physics providing and solution to $\Delta a_{\mu}$ from $\sim 14$ TeV to $\sim 8$ TeV.

We show the corresponding limits derived in the SSF models in the bottom row. The behavior of the solid and dashed contours follows the same logic as in Fig.~\ref{fig:scales2}. However, as mentioned there is no analogous region on nonperturbativity for SSF models based on the overall size of couplings. One approach to evaluate the perturbativity limit of these models is to demand unitarity of $2\to 2$ scalar scattering amplitudes in a given model~\cite{Capdevilla:2021rwo,Goodsell:2018tti}. However, this bound highly depends on the values of scalar quartic couplings in the model which are beyond our considerations. Rather, we restrict the values of this coupling such that no particles in the scalar spectrum of the model become tachyonic. For SSF models the dotted lines show the bound from this requirement.

%
%
\section{Discussion} 
\label{sec:discussion}
In the previous sections, we derived the so-called $k$ equation and explored its implications in the context of minimal models for mass-enhanced corrections to the muon anomalous magnetic moment. Here we elaborate on possible features that could appear in non-minimal models or including sub-dominant contributions and the expected corrections to the $k$-equation.

Tree models with heavy leptons where the SM Higgs acts as only a single component of an extended Higgs sector participating in EWSB present only a minor modification to our arguments. In this case, the $k$ equation will generically be parameterized by an additional free parameter related to the mixing in the Higgs sector. For example, in the case of a 2HDM type-II, this is determined by the ratio of vacuum expectation values of the two Higgs doublets, $\tan\beta$, see~\cite{Dermisek:2021ajd,Dermisek:2021mhi}, and the modification to Eq.~\ref{eq:tree_X_SM}, assuming that new leptons are the heaviest particles in the spectrum, becomes
\begin{equation}
k=\frac{64\pi^{2}}{\mathcal{Q}(1+\tan^{2}\beta)}.
\label{eq:tree_X_2HDM}
\end{equation}
For further a more in depth study of this case, including the pattern of corrections assuming arbitrary Higgs masses compared to new leptons and corrections from scalar quartic couplings, see~\cite{Dermisek:23}.

Another example, which is more generic, arises when considering sub-leading contributions to $C_{\mu\gamma}$. For instance, in the tree models corrections of order $(v/M)^{2}$ appear in the mass eigenstate basis from loops of gauge bosons where a mixing angle between light and heavy leptons is necessary to generate the diagram. In this case, we have 
\begin{equation}
C_{\mu\gamma} \simeq \frac{k}{e}C_{\mu H} + \Delta.
\end{equation}
Note that if $\phi(\Delta)\neq\phi(C_{\mu H})$ then the overall phase of $C_{\mu\gamma}$ will be different than $C_{\mu H}$. This leftover phase, denoted by $\phi_{k}$ in Eq.~\ref{eq:ellipse}, has the effect of shifting the center of the ellipse. The same comment would apply to any non-minimal model with additional fields and couplings which could generate a $\Delta$-like term. For example, in tree models when $C_{\mu\gamma}$, and hence the muon mass, is generated by two sources of chiral enhancement we have
\begin{flalign}
C_{\mu\gamma}=C_{\mu\gamma}^{A}+C_{\mu\gamma}^{B}=\frac{k^{A}}{e}C_{\mu H}^{A} + \frac{k^{B}}{e}C_{\mu H}^{B},
\label{eq:AB_k}
\end{flalign}
where $C_{\mu\gamma}^{A,B}$ and $C_{\mu H}^{A,B}$ denote two independent corrections to either Wilson coefficient. We may then define $k$ by
\begin{flalign}
k^{AB}\equiv (k^{A}C_{\mu H}^{A} + k^{B}C_{\mu H}^{B})/C_{\mu H},
\end{flalign}
where $C_{\mu H}=C_{\mu H}^{A} + C_{\mu H}^{B}$. Then, we have
\begin{flalign}
C_{\mu\gamma} = \frac{k^{AB}}{e}C_{\mu H}.
\label{eq:gen_k}
\end{flalign}
Thus, clearly if the phases $\phi\left(C_{\mu H}^{A}\right)\neq \phi\left(C_{\mu H}^{B}\right)$ then the phase of $C_{\mu\gamma}$ will be different than $C_{\mu H}$. This occurs, for instance, also in the tree models from the subdominant contribution proportional to $y_{\mu}$, if $y_{\mu} \in \mathbb{C}$. Eq.~\ref{eq:gen_k} holds also for loop models where multiple couplings may generate a chiral enhancement simultaneously, e.g. $\lambda_{YZ}$ and $\lambda^{\prime}_{YZ}$ in FFS models. In this case, $k^{AB}$ may not be separable as in Eq.~\ref{eq:AB_k} and will generally be complex, see the Appendix.

In these examples we see that generically $k\in \mathbb{C}$ when including additional corrections in non-minimal models or even from subdominant contributions. The effect of this, is to either stretch the ellipse, when $|k|$ is modified, or shifts the center of the ellipse, when $\phi_{k}$ is modified, or both. 

As a final consideration, we have noted that scalar quartic couplings play a role in both FFS- and SSF-type models. This can also distort the ellipse in the ways we have described for $\lambda_{\phi}\in \mathbb{C}$ in Eq.~\ref{eq:loop_k}. Additionally, scalar quartic couplings can play a role in tree-models with heavy Higgs bosons.

\subsection{RG effects}
\label{sec:RG}
Up to this point our discussion has focused on correlations of dimension-six operators obtained at the matching scale $\Lambda$. Generally speaking, solutions to the measured discrepancy in $\Delta a_{\mu}$ via a chiral enhancement are attractive as they can generate the needed correction assuming perturbative couplings with a mass scale of new physics $\Lambda \gg 1$ TeV, and thus can easily evade direct and indirect constraints. However, with an increasing mass scale of new physics or even a mild scale with a range of couplings up to the limit of perturbativity it is expected that RG induced effects also become increasingly important. We turn to a discussion of these effects in this section. We have followed the conventions of~\cite{Jenkins:2013zja,Jenkins:2013wua,Alonso:2013hga} where, in particular, $\dot{C}_{i}=16\pi^{2}\mu\;dC/d\mu$.

Wilson coefficients calculated at the matching scale, $\Lambda$, are evolved to lower energies via the renormalization group. For $\Lambda$ not too far above the weak scale (justifying the leading log expression), Wilson coefficients at $m_{Z}$ are given by 
\begin{equation}
C_{i}(m_{Z})\simeq C_{i}(\Lambda) - \frac{1}{16\pi^{2}}\gamma_{ij}C_{j}(\Lambda)\log\left(\Lambda/m_{Z}\right),
\label{eq:RG}
\end{equation}
where $\gamma_{ij}$ is the anomalous dimension leading to the renormalization of $C_{i}$ due to $C_{j}$. The complete set of renormalization group equations for SMEFT in the Warsaw basis is given in~\cite{Jenkins:2013zja,Jenkins:2013wua,Alonso:2013hga}.

The RG evolution of lepton dipole operators has recently been discussed extensively in~\cite{Aebischer:2021uvt}. There it was pointed out that for a large enough scale of new physics RG mixing of the four-fermion operator $C^{(3)}_{lequ}\epsilon_{jk}(\bar{l}_{L})^{j}\sigma_{\mu\nu}\mu_{R}(\bar{q})^{k}\sigma^{\mu\nu}u_{R}$ (contraction is with respect to doublet components) with the dipole operators can be competitive with the tree level term in Eq.~\ref{eq:RG}. This effect is then expected to be relevant when $C^{(3)}_{lequ}$ is generated at tree level at the matching scale, such as what occurs in scalar leptoquark models~\cite{deBlas:2017xtg,Feruglio:2018fxo,Gherardi:2020det,Aebischer:2021uvt,Dedes:2021abc}.

With this is mind, the question of radiative stability of Eq.~\ref{eq:WC_relation} comes into play. Retaining only effects driven by the SM top Yukawa coupling the relevant RGE's simplify to
\begin{flalign}
\dot{C}_{\mu H}&\simeq 4N_{C}y_{t}^{3}C^{(1)}_{lequ} + 3N_{C}y_{t}^{2}C_{\mu H},\\
\dot{C}_{\mu\gamma} &\simeq \frac{10}{3}eN_{C}y_{t}C^{(3)}_{lequ} + N_{C}y_{t}^{2}C_{\mu\gamma},
\end{flalign}
where $N_{C}=3$ is the number of colors for SM quarks.
In the absence of four-fermion operators, we see that the RG effects for $C_{\mu H}$ and $C_{\mu \gamma}$ are simply due to themselves and no operator mixing is present in this limit. It is clear, then, that the RG equation for $k$ as defined by Eq.~\ref{eq:WC_relation} will also be proportional to itself and we find a simple approximate solution for the running of this parameter

\begin{equation}
k(\mu)\simeq k(\Lambda)\left(\frac{\Lambda}{\mu}\right)^{-\frac{N_{C}y_{t}^{2}}{8\pi^{2}}}.
\end{equation}
Thus, for new physics at $\Lambda = 1 - 10$ TeV, $k(m_{Z})$ receives RG corrections of $\sim10\%$.

In the opposite limit, where $C^{(1)}_{lequ}$ and $C^{(3)}_{lequ}$ dominate the RG evolution we have 
\begin{equation}
\dot{k}\simeq \frac{eN_{C}y_{t}}{C_{\mu\gamma}}\left(4y_{t}^{2}C^{(1)}_{lequ}-\frac{10}{3}kC^{(3)}_{lequ}\right).
\label{eq:k_RG_FF}
\end{equation}
Schematically, we see that in models where $C^{(1)}_{lequ}$ and $C^{(3)}_{lequ}$ are generated at tree level the running of Eq.~\ref{eq:WC_relation} can be large, $\dot{k}\sim 16\pi^{2}$. It is worth mentioning that in the scalar leptoquark models with a single new particle $C^{(1)}_{lequ} \propto C^{(3)}_{lequ}$ (see, for example, the models discussed in~\cite{Aebischer:2021uvt}). Further, the chirally-enhanced contributions to $C_{\mu\gamma}$ are generated by the same couplings and  $C^{(1)}_{lequ} \propto C^{(3)}_{lequ} \propto C_{\mu\gamma}$. Thus, the RG evolution defined by Eq.~\ref{eq:k_RG_FF} generates large corrections to $k$, but is determined solely by SM couplings.
%
%
\section{Conclusions} 
\label{sec:conclusions}
In this paper, we have argued that in almost all classes of models which generate mass-enhanced corrections to the muon anomalous magnetic moment, the effective operator which generates the muon dipole moment is correlated with the operator which generates a correction to the muon mass, where the latter manifests as deviations from the SM prediction for $h\to\mu^{+}\mu^{-}$. At the matching scale, these operators are dictated by couplings of the SM Higgs to new fermions or new scalars. Thus, the resulting low-energy predictions are sensitive probes of dynamics of Higgs couplings beyond the SM.

The correlation between the dipole operators with corrections to the muon mass leads to a correlation between the ratio of $h\to\mu^{+}\mu^{-}$ compared to that in the SM, $R_{h\to\mu^{+}\mu^{-}}$, and the electric and magnetic dipole moments of the muon, $\Delta a_{\mu}$ and $d_{\mu}$, which we refer to as the $k$ equation. Future measurements of $R_{h\to\mu^{+}\mu^{-}}$ are expected to reach a precision of $\mathcal{O}(1)\%$ within the SM prediction. The correlation to $\Delta a_{\mu}$, assuming that the measured deviation persists, leads to predictions of $d_{\mu}$ which in many models are within reach of upcoming experiments. 

Our study shows that in the context of the minimal models we have considered, the pattern of deviation defined by the muon ellipse allows to set upper limits on the scale of new physics that can be stronger than more general considerations such as perturbative unitarity or tachyonic particles. In particular, we find that for FFS- and SSF-type models which can explain the central value of $\Delta a_{\mu}$, future measurements of $R_{h\to \mu^{+}\mu^{-}}$ reaching 10\% to 1\%  can reduce the maximum mass scale of new physics by close to a factor of two within this projected range of precision, conservatively assuming that the limit on $|d_{\mu}|$ is probed at least to the FNAL projection. Thus, we advocate that within the class solutions to $\Delta a_{\mu}$ involving a chirally-enhanced dipole moment the correlation of the predictions of $R_{h\to \mu^{+}\mu^{-}}$ and $|d_{\mu}|$ defined by the muon ellipse can have a significant impact on the allowed parameter space in a given model.

For simplicity and breadth, we have restricted our discussion to simplified models which contain the minimal particle content to generate mass-enhanced corrections to $\Delta a_{\mu}$. However, the $k$ equation has applications in a variety of well-motivated UV completions. Vectorlike leptons are commonplace in both non-supersymmetric and supersymmetric GUT's, the spectrum of the FFS models are prototypical in extensions of the SM with scalar leptoquarks (where the fermions in the loop correspond to the left- and right-handed top quarks), and the spectrum and corresponding corrections in the SSF type models are analogous to SUSY corrections arising from smuon-Bino loops in the minimal supersymmetric SM. Going beyond these examples, having discussed the implications for models with $SU(2)$ triplets or more generic hypercharges and bridge-type models, essentially covers an infinite class of possible models with phenomenological implications of the $k$ equation.

It is worth noting that in this work we have elaborated on the implications of the correlation provided by the $k$ equation relying minimally on the details of the models which are needed to generate chirally-enhanced corrections to $C_{\mu\gamma}$ and $C_{\mu H}$. In a more complete setting other constraints may come in to play. For example, in loop models we have largely ignored details of the scalar potential and, as we have commented on, this may provide further constraints with respect to perturbative unitarity. Similarly, some models may also generate the dimension-5 operator $l_{L}l_{L}HH$ providing a connection to the neutrino sector of the SM and associated phenomenology. Though, these and similar considerations are not generic to the full classes of models we have discussed.

Many puzzles remain regarding the second generation SM fermions. In the near future, the couplings of the muon in particular will be intensely scrutinized. In this work, we have outlined a novel way to consider corrections to $\Delta a_{\mu}$, $R_{h\to\mu^{+}\mu^{-}}$, and $d_{\mu}$ from new physics, informing the pattern of deviations expected in the presence of a signal.

\acknowledgments
The work of R.D. was supported in part by the U.S. Department of Energy under Award No. {DE}-SC0010120. TRIUMF receives federal funding via a contribution agreement with the National Research Council of Canada.

\appendix
\section{Contributions in FFS-type Models}
In the following appendices, we present the matching of FFS-type models to $C_{\mu B,W}$ and $C_{\mu H}$ based on Eq.~\ref{eq:FFS_lag}. We consider models with $SU(2)$ doublets and triplets discussed in~\cite{Crivellin:2021rbq}, though we present corrections proportional to $\lambda_{YZ}$ and $\lambda^{\prime}_{YZ}$ separately.

One loop corrections to $C_{\mu B}$ and  $C_{\mu W}$ based on Eq.~\ref{eq:FFS_lag} are given by
\begin{flalign}
    \begin{split}
        C_{\mu B} = - \left(\frac{\lambda_Y \lambda_Z \lambda_{YZ}}{64 \pi^2 M_Y M_Z} \right) g^{\prime} & \left[Y_Y (A(x_Y, x_Z) + B(x_Y, x_Z)) \right. \\
        & \left. + Y_Z (A(x_Z, x_Y) + B(x_Z, x_Y)) - 2 Y_\Phi C(x_Y, x_Z) \right] \xi_{eB}\\
        - \left(\frac{\lambda_Y \lambda_Z \lambda_{YZ}'}{64 \pi^2 M_Y M_Z} \right) g^{\prime} & \left[Y_Y \left(2 I(x_Y, x_Z) + \left(\frac{M_Y}{M_Z} \right) A(x_Y,x_Z) \right) \right. \\ 
        & \left. + Y_Z \left(2 I(x_Z, x_Y) + \left(\frac{M_Z}{M_Y} \right) A(x_Z,x_Y) \right) + 2 Y_{\Phi} J(x_Y, x_Z) \right] \xi_{eB},
    \end{split}
\end{flalign}
and
\begin{flalign}
    \begin{split}
        C_{\mu W} = - \left(\frac{\lambda_Y \lambda_Z \lambda_{YZ}}{64 \pi^2 M_Y M_Z} \right) g & \left[\xi_Y (A(x_Y, x_Z) + B(x_Y, x_Z)) \right. \\ 
        & \left. + \xi_Z (A(x_Z, x_Y) + B(x_Z, x_Y)) - 2 \xi_\Phi C(x_Y, x_Z) \right]\\
        - \left(\frac{\lambda_Y \lambda_Z \lambda_{YZ}'}{64 \pi^2 M_Y M_Z} \right) g & \left[ \xi_Y \left(2 I(x_Y, x_Z) + \left(\frac{M_Y}{M_Z} \right) A(x_Y,x_Z) \right) \right. \\ 
        & \left. + \xi_Z \left(2 I(x_Z, x_Y) + \left(\frac{M_Z}{M_Y} \right) A(x_Z,x_Y) \right) + 2\xi_{\Phi} J(x_Y, x_Z) \right],
    \end{split}
\end{flalign}
\begin{table}[t]
\begin{center}
\begin{tabular}{ |c||c|c|c|c|c| } 
\hline
$(Y,Z,\Phi)$ & $\xi_Y$ & $\xi_Z$ & $\xi_{\Phi}$ & $\xi_{eB}$ & $\xi_{eH}$ \\
\hline
$121$ &$0$& $1/2$ & $0$ & $1$ & $1$\\
\hline
$212$  & $1/2$ & $0$ & $-1/2$ & $-1$ & $-1$\\
\hline
$323$ &$-2$ & $ -1/2$ & $2$ & $3$ & $5$\\
\hline
$232$ & $1/2$ & $2$ & $-1/2$ & $3$ & $5$\\
\hline
\end{tabular}
\caption{$SU(2)$ group theoretical $\xi_i$ factors entering the contributions to Wilson coefficients $C_{\mu B}, C_{\mu W}$ and $C_{\mu H}$ for representations of the $(Y,Z,\Phi)$ fields in Eq.~\ref{eq:FFS_lag}.}
\label{table:xi_factors}
\end{center}
\end{table}
where $Y_{Y,Z}$ and $Y_{\Phi}$ are the hypercharge numbers of the new fermions and scalar, respectively, the representation-dependent $\xi_i$ factors are defined in Table~\ref{table:xi_factors}, and $x_{i}=M_{i}^{2}/M_{\Phi}^{2}$ for $i=Y,Z$. Formulae for the loop functions are collected in Appendix B. The contributions to $C_{\mu H}$ are
\begin{flalign}
    \begin{split}
    C_{\mu H} = - & \left( \frac{\lambda_Y \lambda_Z}{16 \pi^2 M_{Y}M_{Z}} \right) \left[ |\lambda_{YZ}|^2 \lambda_{YZ} K(x_Y, x_Z) - |\lambda_{YZ}'|^2 \lambda_{YZ}' L(x_Y, x_Z) - (\lambda_{YZ})^2 \lambda_{YZ}'^{*} M(x_Y, x_Z) \right. \\ \\
    & \left. + \left( |\lambda_{YZ}|^2 \lambda_{YZ}' \left(\frac{M_Y}{M_Z} + \frac{M_Z}{M_Y} \right) + 2 |\lambda_{YZ}'|^2 \lambda_{YZ} + (\lambda_{YZ}')^2 \lambda_{YZ}^* \right) K(x_Y, x_Z) \right] \xi_{eH}.
    \end{split}
\end{flalign}
Note that for a given set of masses of new particles, these corrections lead to a relation
\begin{flalign}
C_{\mu\gamma}=\frac{k(\lambda_{YZ},\lambda^{\prime}_{YZ})}{e}C_{\mu H},
\end{flalign}
such that the $k$ equation defines a two-parameter family of ellipses for $\Delta a_{\mu},~d_{\mu}$ and $R_{h\to\mu^{+}\mu^{-}}$ parameterized by the values of $\lambda_{YZ}$ and $\lambda^{\prime}_{YZ}$. It is clear in this case that $k\in \mathbb{C}$.

In the equal-mass limit, $M_Y = M_Z = M_{\Phi}$, when $\lambda^{\prime}_{YZ}=0$ we obtain
\begin{flalign}
    C_{\mu \gamma} =  - \left(\frac{\lambda_Y \lambda_Z \lambda_{YZ}}{384 \pi^2 M^2} \right) e \left[ \left( \frac{3}{2} \left(Y_Y + Y_Z \right) - Y_{\Phi} \right)\xi_{eB} - \left(\frac{3}{2} \left(\xi_Y + \xi_Z \right) - \xi_{\Phi} \right) \right],
\end{flalign}
and 
\begin{flalign}
    C_{\mu H} = - \left(\frac{\lambda_Y \lambda_Z \lambda_{YZ}}{192 \pi^2 M^2} \right) |\lambda_{YZ}|^2 \xi_{eH}.
\end{flalign}
Thus, the corresponding $\mathcal{Q}$ factor defined by Eq.~\ref{eq:loop_Qs1} is given by
\begin{flalign}
    \begin{split}
        \mathcal{Q}_{FFS} \xi_{eH} & = 2 \left[ \left(\frac{3}{2} \left(Y_Y + Y_Z \right) - Y_{\Phi} \right) \xi_{eB} - \left( \frac{3}{2} \left(\xi_Y + \xi_Z \right) -\xi_{\Phi} \right)  \right], \\
        & = -2 \left[ \left(\frac{9}{4} + 4 Y_{\Phi} \right) \xi_{eB} + \frac{3}{2} \left(\xi_Y + \xi_Z \right) - \xi_{\Phi} \right],
    \end{split}
\label{eq:Q_lambda}
\end{flalign}
where we used $1/2 + Y_Z + Y_{\Phi} = 0$ and $1 + Y_Y + Y_{\Phi} = 0$ by conservation of hypercharge. Similarly, in the absence of $\lambda_{YZ}$ we find 

\begin{flalign}
    C_{\mu \gamma} = \left(\frac{\lambda_Y \lambda_Z \lambda_{YZ}'}{768 \pi^2 M^2} \right) e \left[ \left(Y_Y + Y_Z - 2Y_{\Phi} \right)\xi_{eB} - \left( \xi_Y + \xi_Z - 2 \xi_{\Phi} \right) \right],
\end{flalign}
and 
\begin{flalign}
    C_{\mu H} = \left( \frac{3 \lambda_Y \lambda_Z \lambda_{YZ}'}{192 \pi^2 M^2} \right)|\lambda_{YZ}'|^2 \xi_{eH},
\end{flalign}
and the corresponding $\mathcal{Q}$ factor is given by
\begin{flalign}
    \begin{split}
        \mathcal{Q}'_{FFS} \xi_{eH} & = \frac{1}{3} \left[(Y_Y + Y_Z - 2Y_{\Phi})\xi_{eB} - (\xi_Y + \xi_Z - 2 \xi_{\Phi}) \right] \\
        & = - \frac{1}{3} \left[\left( \frac{3}{2} + 4Y_{\Phi}\right) \xi_{eB} + \xi_Y + \xi_Z - 2 \xi_{\Phi} \right].
    \end{split}
\label{eq:Q_lambda_prime}
\end{flalign}
Finally, considering models when $\lambda_{YZ}' = \lambda_{YZ}$, we find

\begin{flalign}
    C_{\mu \gamma} = -\left(\frac{\lambda_Y \lambda_Z \lambda_{YZ}}{384 \pi^2 M^2} \right) e \left[ \left(Y_Y + Y_Z \right) \xi_{eB} - \left( \xi_Y + \xi_Z \right) \right],
\end{flalign}
and 
\begin{flalign}
    C_{\mu H} = - \left( \frac{\lambda_Y \lambda_Z \lambda_{YZ}}{96 \pi^2 M^2} \right)|\lambda_{YZ}|^2 \xi_{eH},
\end{flalign}
with a $\mathcal{Q}$ factor given by
\begin{flalign}
    \begin{split}
        \mathcal{Q}_{FFS}^{(\lambda_{YZ}=\lambda_{YZ}^{\prime})} \xi_{eH} & = \left[(Y_Y + Y_Z) \xi_{eB} - (\xi_Y + \xi_Z) \right] \\
        & = - \left[\left(\frac{3}{2} + 2Y_{\Phi} \right) \xi_{eB} + \xi_Y + \xi_Z \right].
    \end{split}
\label{eq:Q_total}
\end{flalign}

\section{Loop Functions}
The loop functions relevant for FFS-type models are given by  
\begin{flalign}
         A(x,y) &= \frac{xy}{2} \left[\frac{-3y + x(1+x+y)}{(x-1)^2 (x-y)^2} + \frac{2 (x^3 + x^2 y (x-3) + y^2) \ \textrm{ln}(x)}{(1-x)^3(x-y)^3} - \frac{2 y^2 \ \textrm{ln}(y)}{(x-y)^3 (1-y)} \right], \\
 B(x,y) &= \frac{xy}{2(x-y)} \left[\frac{(1-y)(y-3) - 2 \ \textrm{ln}(y)}{(1-y)^3} - \frac{(1-x)(x-3) - 2 \ \textrm{ln}(x)}{(1-x)^3} \right] - A(y,x), \\
C(x,y) &= \frac{xy}{2} \left[ \frac{x + xy + y - 3}{(1-x)^2(1-y)^2} - \frac{2 x \ \textrm{ln}(x)}{(x-y)(1-x)^3} + \frac{2 y \ \textrm{ln}(y)}{(x-y)(1-y)^3} \right],\\\nonumber
I(x,y) & = \frac{\sqrt{xy}}{12} \left[ \frac{3\left(x^2(3-x) + xy(x-3)(1+x) + y^2(2+x(x-1)) \right)}{(1-x)^2(x-y)^2(1-y)} \right. \\\nonumber
        & \left. + \frac{2y(4x^2-2xy+y^2) \ \textrm{ln}(x/y)}{(x-y)^4} \right. \\ \nonumber
        & \left. +  \frac{2\left(3x^4+x^2y(x^2(x-6) - 4) + xy^2(2+6x + x^3) + y^3(x^2(x-3)-1)\right) \textrm{ln}(x)}{(1-x)^3 (x-y)^4} \right. \\
        & \left. + \frac{2y\left(xy(y(y-5)-2) + x^2(4+y(y-2)) + y^2 (1+y+y^2) \right) \textrm{ln}(y)}{(1-y)^2(x-y)^4}\right], \\
 J(x,y) &= \frac{\sqrt{xy}}{2} 
        \left[ \frac{1+x+y-3xy}{(1 -x)^2(1-y)^2} + \frac{2x^2 \ \textrm{ln}(x)}{(1-x)^3(x-y)} - \frac{2y^2 \ \textrm{ln}(y)}{(1-y)^3 (x-y)}\right],\\
K(x,y) &= xy \left[ \frac{x + y -2xy}{(1-x)(1-y)(x-y)^2} + \frac{x(x^2 + y(x-2)) \ \textrm{ln}(x)}{(1-x)^2(x-y)^3} - \frac{y(y^2 + x(y-2)) \ \textrm{ln}(y)}{(1-y)^2(x-y)^3} \right], \\
L(x,y) &= \sqrt{xy} \left[\frac{x^2y+y^2x - x^2 - y^2 }{(1-x)(1-y)(x-y)^2} - \frac{x^2(x-3y+2xy) \ \textrm{ln}(x)}{(1-x)^2(x-y)^3} + \frac{y^2(y-3x+2xy) \ \textrm{ln}(y)}{(1-y)^2(x-y)^3} \right], \\ 
M(x,y) &= (xy)^{3/2} \left[ \frac{x+y-2}{(1-x)(1-y)(x-y)^2} + \frac{(x-2x^2+y) \ \textrm{ln}(x)}{(1-x)^2(x-y)^3} - \frac{(y-2y^2+x) \ \textrm{ln}(y)}{(1-y)^2(x-y)^3}\right], 
\end{flalign}
Note that functions $A(x,y),~B(x,y)$, and $C(x,y)$ are the same as those appearing in the matching calculations for tree models. Useful limits for approximate formula found in the text are
\begin{flalign}
\lim_{x, y \rightarrow \infty} A(x,y) = \frac{1}{6}, \ \ \ \ \ \ \ \ \ \lim_{x,  y \rightarrow 1} A(x,y) = \frac{1}{12},
\end{flalign}
\begin{flalign}
\lim_{x, y \rightarrow \infty} B(x,y) = \frac{1}{3}, \ \ \ \ \ \ \ \ \ \lim_{x,  y \rightarrow 1} B(x,y) = \frac{1}{6},
\end{flalign}
\begin{flalign}
\lim_{x, y \rightarrow \infty} C(x,y) = \frac{1}{2}, \ \ \ \ \ \ \ \ \ \lim_{x, y \rightarrow 1} C(x,y) = \frac{1}{12},
\end{flalign}
\begin{flalign}
 \lim_{x, y \rightarrow \infty} I(x,y) = -\frac{1}{12}, \ \ \ \ \ \ \ \ \ \lim_{x, y \rightarrow 1} I(x,y) = -\frac{1}{12},
\end{flalign}
\begin{flalign}
\lim_{x, y \rightarrow \infty} J(x,y) = 0, \ \ \ \ \ \ \ \ \ \lim_{x, y \rightarrow 1} J(x,y) = \frac{1}{12}, 
\end{flalign}
\begin{flalign}
\lim_{x, y \rightarrow \infty} K(x,y) = \frac{1}{6}, \ \ \ \ \ \ \ \ \ \lim_{x, y \rightarrow 1} K(x,y) = \frac{1}{12},
\end{flalign}
\begin{flalign}
\lim_{x, y \rightarrow \infty} L(x,y) = \frac{1}{3}, \ \ \ \ \ \ \ \ \ \lim_{x, y \rightarrow 1} L(x,y) = \frac{1}{4},
\end{flalign}
\begin{flalign}
\lim_{x, y \rightarrow \infty} M(x,y) = \frac{1}{3}, \ \ \ \ \ \ \ \ \ \lim_{x, y \rightarrow 1} M(x,y) = \frac{1}{12}.
\end{flalign}


\bibliography{ref}

\clearpage
\newpage






\end{document}